\documentclass[useAMS,usenatbib]{mn2e}
\usepackage{txfonts,graphicx,amssymb,subfigure}
\usepackage[normalem]{ulem}
\usepackage{natbib}

\begin{document}

\title[The cosmic-ray electron population in NGC~3801]{
The impact of a young radio galaxy: clues from the cosmic-ray
electron population}

\author[V.~Heesen, J.~H.~Croston, J.~J.~Harwood, M.~J.~Hardcastle and
A.~Hota]{Volker Heesen$^{1}$\thanks{E-mail: v.heesen@soton.ac.uk}, Judith
  H.~Croston$^{1}$, Jeremy J.~Harwood$^{2}$, Martin J.~Hardcastle$^{2}$ \newauthor and Ananda Hota$^{3}$\\
$^{1}$School of Physics and Astronomy, University of Southampton,
Southampton SO17 1BJ, UK\\
$^{2}$School of Physics, Astronomy and Mathematics, University of
Hertfordshire, Hatfield AL10 9AB, UK\\
$^{3}$UM-DAE Centre for Excellence in Basic Sciences, Vidyanagari, Mumbai
400098, India}

\voffset-.6in
 
\date{Accepted 2014 January 7. Received 2013 December 21; in original form 2013 October 4}

\maketitle

\begin{abstract}
  In the framework of hierarchical structure formation AGN feedback shapes the
  galaxy luminosity function. Low luminosity, galaxy-scale double radio
  sources are ideal targets to investigate the interplay between AGN feedback
  and star formation. We use VLA and BIMA observations to study the radio
  continuum emission of NGC~3801 between $1.4$ and $112.4~\rm GHz$. We find a
  prominent spectral break at $\approx$10~GHz, where the spectrum steepens as
  expected from cosmic-ray electron (CRe) ageing. Using the equipartition
  magnetic field and fitting JP models locally we create a spatially resolved
  map of the spectral age of the CRe population. The spectral age of
  $\tau_{\rm int}=2.0\pm 0.2~\rm Myr$ agrees within a factor of two with the
  dynamical age of the expanding X-ray emitting shells. The spectral age
  varies only little across the lobes, requiring an effective mixing process
  of the CRe such as a convective backflow of magnetized plasma. The jet
  termination points have a slightly younger CRe spectral age, hinting at
  in-situ CRe re-acceleration. Our findings support the scenario where the
  supersonically expanding radio lobes heat the ISM of NGC~3801 via shock
  waves, and, as their energy is comparable to the energy of the ISM, are
  clearly able to influence the galaxy's further evolution.
\end{abstract}

\begin{keywords}
cosmic rays -- galaxies: individual: NGC~3801 -- galaxies: quasars: general --
galaxies: jets -- radiation mechanisms: non-thermal -- radio continuum: galaxies
\end{keywords}
  
\section{Introduction}
In the framework of the $\Lambda$CDM Universe hierarchical structure formation
feedback, either by star formation or processes connected to active galactic
nuclei (AGN), shape the observed galaxy population in today's Universe
\citep[e.g.,][]{croton_06a}. Powerful FR~I and FR~II radio galaxies are well studied but
relatively rare; the energy input from lower luminosity jets may be as
important for galaxy evolution, but it is much less well
understood. Low-luminosity, galaxy-scale double radio
sources on scales of 1--10~kpc are found in a number of elliptical
galaxies, where they either resemble young, twin-jet FR~I sources \citep[e.g.,
NGC~1052, NGC~3801,][]{kadler_04a,croston_07a} or in Seyfert and spiral
galaxies where they typically show twin bubble/lobe features without obvious
jets \citep[NGC~6764, Markarian~6,][]{croston_08a,mingo_11a}. In both cases
the jet-driven radio outflows can inject large amounts of energy into their
environments, and may represent an important galaxy feedback mechanism
\citep{croston_08b}.
Shock heating by the radio lobes is a very efficient way of transferring
energy from the AGN to the surrounding galaxy environment . The best studied
example is Centaurus~A, a low-power FR~I source, where $\rm Mach \approx 8$
shocks surrounding the inner radio lobes heat the gas and accelerate cosmic
rays to high energies \citep{kraft_03a, croston_09a}. Deep $\emph Chandra$
observations show the X-ray emitting gas to be concentrated in shells that are
upstream of the radio lobes, constraining the jump conditions in the
interstellar medium (ISM). The energy content of the radio lobes is a few
times $10^{55}~\rm erg$, comparable to the estimated thermal energy of the
host galaxy's ISM. This can be seen when one multiplies the energy
  density of the ISM, $3.8\times 10^{-12}~\rm erg\,cm^{-3}$
  \citep{croston_07a}, with the volume within 11~kpc radius assuming a
  vertical height of 100~pc on both sides of the galactic midplane, which
  results in a thermal energy of $9\times 10^{55}~\rm erg$. A number of
additional galaxies with supersonically expanding lobes have been found
including NGC~3801 \citep{croston_07a}, Markarian~6 \citep{mingo_11a} and the
Circinus galaxy \citep{mingo_12a}. The lobes of large FR~I radio galaxies are
not thought to be over-pressurized, but the majority of sources are likely to
pass through a stage of evolution in which the lobes inflate supersonically
and can hence drive shocks into their environment \citep[e.g.,][]{heinz_98a}.
For understanding the dynamics and evolution of radio galaxies, it is
important to establish the particle content in FR~I jets and lobes. For the
powerful FR~II lobes, the synchrotron minimum-energy estimates are roughly
in equipartition with the external pressure measurements
\citep[e.g.,][]{hardcastle_02a,belsole_04a,croston_04a}. This finding is
corroborated by inverse Compton (IC) X-ray emission, which indicates that the
electron energy density in FR~II lobes are almost in equipartition with the
magnetic field energy density in the absence of an energetically important
proton population \citep[e.g.,][]{croston_05a,kataoka_05a}. On the other hand,
it has been known for a long time that in lower-power (FR~I) radio galaxies
the cosmic-ray electrons (CRe) as derived from energy equipartition can not
provide enough pressure to balance that of the external medium
\citep[e.g.,][]{morganti_88a,hardcastle_98a,worrall_00a}. Electrons can not be
responsible for the missing energy as they would be detected in X-ray emission
via IC radiation which is not observed \citep{croston_03a}. Hot,
thermal gas with a temperature similar to that of the external medium would
also be detectable in X-ray emission. \citet{croston_08b} showed that entrainment
likely plays an important role, with protons contributing the missing
pressure: so-called `bridged' FR~I radio galaxies, where the jet has only a
small contact area with the environment, have pressures much closer to the
equipartition value than `plumed' FR~I sources, where the jet seemingly
interacts with the ISM. A detailed study of the particle content in the plumed
FR~I source 3C~31 further supports this scenario \citep{croston_14a}.
Radio continuum (RC) observations are a good way to establish the content of
the relativistic cosmic rays and magnetic fields. This is particularly the
case if one has a wide range of frequencies available to quantify the spectrum
of the RC emission which can be used in conjunction with electron ageing
models \citep[e.g.,][]{alexander_87a}. Single-burst injection models
\citep[e.g.,][]{jaffe_73a} have been shown to provide good fits to the
spectral behaviour of lobes in large radio galaxies
\citep[e.g.,][]{alexander_87b,carilli_91a,liu_92a,hardcastle_08a,harwood_13a}. They
predict a characteristic break frequency in the synchrotron spectrum of the
aged plasma, which is related to the magnetic field strength and the time
since the population was accelerated. The spectral age estimate can be
compared with the dynamical time-scales derived from the X-ray emission and
the lobe expansion velocity and lets us investigate the history of the
particle acceleration in the lobes. This is crucial for determining the energy
released in shocks throughout the life of the source. In addition, this
technique provides us with information about the spatial distribution of the
CRe age, something that can be used to locate the areas where the young plasma
is injected into the lobes \citep{harwood_13a}. There is some discussion as to
the extent to which spectral ages are reliable, because in some radio galaxies
the spectral and dynamical age as obtained from the source expansion history
are discrepant
\citep[e.g.,][]{rudnick_94a,eilek_96a,blundell_00a,hardcastle_05a,goodger_08a,harwood_13a}. However,
for young radio galaxies like NGC~3801 with a dynamical age much smaller than
$10^7~\rm yr$, spectral ages have been suggested to be much more accurate
\citep{blundell_00a}.
In this paper, we present new RC observations of the low-power FR~I object
NGC~3801. The radio lobes in this galaxy are small ($\approx 10~\rm kpc$) and
buried within the ISM of the S0/a host galaxy, which shows a disturbed
morphology. It has two prominent dust lanes, one along the minor axis and
another roughly perpendicular extending from the centre to the eastern edge
\citep{heckman_86a}. The S-shaped radio lobes are surrounded by hot X-ray
emitting gas shells detected by \emph{Chandra} observations from which
\citet{croston_07a} derived an expansion velocity of $\rm
850~km\,s^{-1}$. \citet{das_05a}, using the BIMA millimetre-wave array, found
a ring of cool molecular gas with a radius of 2~kpc, aligned with the minor
axis, traced by $^{12}{\rm CO}~J=1\rightarrow 0$ line emission around the
galaxy's nucleus. Atomic hydrogen H\,{\small I} emission imaging studies by
\citet{hota_09a} and subsequently by \citet{emonts_12a} have shown the presence
of a possible gas disc of 30~kpc size and rotating around the host galaxy. The
detection of several compact sources in far-ultraviolet (FUV) radiation using
\emph{GALEX} observations shows star-formation activity in the last few
100~Myr, suggesting that NGC~3801 could be a transition object where the
increased star formation following a gaseous merger is about to be quenched by
the expanding radio lobes that disturb the ISM \citep{hota_12a}.
Throughout this paper, we adopt a cosmology with $H_0=70~\rm
km\,s^{-1}\,Mpc^{-1}$, $\Omega_{\rm M}=0.3$, and $\Omega_\Lambda=0.7$. We
adopt a luminosity distance for NGC~3801 of $52.6~\rm Mpc$, obtained by correcting
the heliocentric velocity of $\rm 3317~km\,s^{-1}$ \citep{lu_93a} to the CMB
frame of reference, which gives an angular scale of $0.25~\rm kpc\,
arcsec^{-1}$. Other properties of NGC~3801 are listed in Table~\ref{tab:NGC3801}.
%
\begin{table}
\caption{General properties of NGC~3801.\label{tab:NGC3801}}
\begin{tabular}{lll}
\hline\hline
Parameter & Value & Reference\\
\hline
Galaxy type & S0/a & 1\\
Galaxy position [$\rm J2000.0$] & $\rm R.A.~11^h40^m16^s.942$ & ... \\
... & $\rm dec.~17\degr 43\arcmin
40\farcs 98$ &  2\\
Other names & UGC~06635, 4C~$+17.52$ & NED \\
Distance & $52.6~\rm Mpc$ & $-$\\
Velocity \& redshift & $3317~\rm km\,s^{-1}$, $z=0.011$ & 3\\
P.A. of dust lane  & $24\degr \pm 2\degr$ & 4\\
P.A. of radio jets & $122\degr\pm 2\degr$ & this paper\\
\hline
\end{tabular}
\medskip\\References -- 1: \citet{huchra_12a}, 2: \citet{evans_10a}, 3: \citet{lu_93a},
4: \citet{verdoes_kleijn_99a}
\end{table}
\section{Observations and data reduction}
\label{sec:observations}
\subsection{VLA observations}
\label{subsec:jansky_vla}
Observations with the Karl G.\ Jansky Very Large Array (VLA)\footnote{The
  National Radio Astronomy Observatory (NRAO) is a facility of the National
  Science Foundation operated under cooperative agreement by Associated
  Universities, Inc.} were taken in March and May 2010 in the `open shared
risk observing' (OSRO) time. We used a bandwidth of 128~MHz with 64 channels
of 2~MHz bandwidth at each observing frequency in $K$-band ($22.0$ and
$25.0~\rm GHz$) and $Ka$-band ($32.3$ and $35.0~\rm GHz$). The sub-band
frequencies in $K$-band and $Ka$-band were both observed simultaneously. We
observed a flux calibrator, 3C~286, once during each observing session, and a
phase calibrator every 5--10~min. For the data reduction we used the
Common Astronomy Software Applications package ({\small CASA})\footnote{Available at {\tt http://casa.nrao.edu}}. The necessary steps are laid out in the continuum tutorial of 3C~391 observations on the CASA homepage\footnote{\tt http://casaguides.nrao.edu/index.php?title=...\\...EVLA\_Continuum\_Tutorial\_3C391}, which we summarize in the following.
We used a model of 3C~286, because the calibrator becomes partially resolved
at our high resolution, and calibrated the flux scale using the
`Perley--Taylor--99' calibration scale \citep{perley_13a}. Our $Ka$-band
observations suffered from a failure to successfully observe the flux
calibrator, 3C~286. This led us carry out some additional steps for the data
reduction as described below. Continuum VLA observations are essentially now
in spectral line mode, so that the bandpass has to be determined first, again
using 3C~286. Phase and amplitude solutions were found by calibrating the secondary
(phase) calibrator J1215+1654 and the flux scale bootstrapped from 3C~286. The
time interpolated solutions were copied to our source NGC~3801.
%
\begin{table*}
\caption{VLA observations of NGC~3801 presented in this paper.\label{tab:VLA}}
\begin{tabular}{rrrrrrrrrrr}
\hline\hline
$\nu$ & Bandwidth & Array & Resolution & Time on & Noise  & Dynamic & Date & Code & References\\
$[\rm GHz]$ & $[\rm MHz]$ &  &  & source & $[\rm \mu Jy\, b.a.^{-1}]$ & range & & &\\
\hline
$1.4$ & 100 & A$+$B & $1.5\times 1.4~{\rm arcsec^2}$ & 1.5$+$0.5~h &$28$ & 1000 & Feb.\ 2006 & AC805 & \citet{croston_07a} \\
$4.9$ & 100 & B$+$C & $1.5\times 1.3~{\rm arcsec^2}$ & 1.0$+$0.5~h &$27$ & 440 & Jun.\ 2006 & AC805 & \citet{croston_07a} \\
$4.9$ & 100 & A & $0.40 \times 0.35~{\rm arcsec^2}$ & 0.8~h &$25$ & 180 & Feb.\ 2006 & AC805 & this paper \\
$22.0$ & 128 & D & $2.7 \times 2.6~{\rm arcsec^2}$ & 1.5~h & $42$ & 240 & Mar.\ 2010 & AC980 & this paper\\
$25.0$ & 128 & D & $2.5 \times 2.4~{\rm arcsec^2}$ & 1.5~h & $32$ & 250 & Mar.\ 2010 & AC980 & this paper\\
$32.3$ & 128 & D & $2.2 \times 1.6~{\rm arcsec^2}$ & 2.8~h & $63$ & 140 & May 2010 & AC980 & this paper\\
$35.0$ & 128 & D & $2.4 \times 1.5~{\rm arcsec^2}$ & 2.8~h & $51$ & 160 & May 2010 & AC980 & this paper\\
\hline
\multicolumn{10}{l}{\emph{Notes.} The observations at $1.4$ and $4.9~\rm GHz$
  were made with the VLA prior to the correlator upgrade. They were also
  combined with pre-existing }\\
\multicolumn{10}{l}{
    archive data (AB920). The $4.9$~GHz data in A-configuration
    were merged with those from the B- and C-configuration.}\\
\end{tabular}
\end{table*}
We checked our data for radio-frequency interference (RFI) prior to
imaging. About $20~\rm per~cent$ of our $K$-band data needed flagging,
either due to RFI or because of weak antenna response resulting in high
antenna noise after calibration, reducing the effective bandwidth to
100~MHz. The theoretical noise level in $K$-band with $1.5~\rm h$ on-source time
are 35 and $30~\mu\rm Jy\,beam^{-1}$ for $22.0$ and $25.0~\rm GHz$,
respectively. Our achieved noise levels are 42 and $30~\mu\rm Jy\,beam^{-1}$
respectively, only 10--20 per cent higher than theoretically expected. In $Ka$-band
we needed to flag 50 and 30 per cent of the data at $32.3$ and $35.0~\rm GHz$,
respectively. This high fraction of data that needed to be flagged is likely
due to a calibration difficulty, which we will discuss below. This reduces the
effective bandwidth to 60 and 90~MHz resulting in theoretical noise levels of
37 and $32~\mu\rm Jy\,beam^{-1}$, respectively, with $2.8~\rm h$ on-source
observing time. Our achieved noise levels of 63 and $51~\mu\rm
Jy\,beam^{-1}$, respectively, are 60--70  per cent above the theoretical
expectation. This can be explained by a generally poor data quality, which was
obvious by inspecting the (u,v)-data visually in {\small CASA}. We notice that the
bandwidth of the observations is relatively small, so that the (u,v)-coverage
is not very much affected by reducing the effective bandwidth.
We self-calibrated the $K$-band data in each channel determining new phase and
amplitude solutions using the task {\small BANDPASS} three times, improving the
r.m.s.\ noise level of our maps. In $Ka$-band (at $35.0~\rm GHz$ only) we
self-calibrated our data in phase only taking all channels together using the
task {\small GAINCAL}. For the imaging we used a multi-frequency multi-scale
(MS--MFS) CLEAN algorithm as described by \citet{rau_11a}. As our available
relative bandwidth was fairly small, it was not necessary to use the spectral index fitting
that this variety of the CLEAN algorithm can handle.
In order to establish an approximate flux scale for our $Ka$-band
observations, we calibrated $Ka$-band data taken by other observing projects
around our observing date of 2010 May 13. We used the averaged flux scale
from May 12 and 14 (both AS1013) to bootstrap the flux density of our
secondary calibrator J1215+1654 as $0.287~\rm Jy$ at $35.0~\rm GHz$. In
$Ka$-band the flux density of this calibrator is known to be changing
significantly on consecutive days, so we calibrated our data with 3C~286
observations on several days. The
observed flux density of our secondary calibrator was ranging from $0.25$ to
$0.35~\rm Jy$, with the exception of May 11 where it was $0.59~\rm
Jy$. Disregarding the outlier on May 11, the flux density scale
variation is $25~\rm per~cent$ which we thus adopt as the systematic error of our flux
density measurements at $Ka$-band. We applied the flux scale as measured at
$35.0~\rm GHz$ also to the observations at $32.3~\rm GHz$, correcting for the
radio spectral index of 3C~286, because at the latter frequency the data was
strongly affected by weak antenna gains, which resulted in a high flagging
fraction of $50~\rm per~cent$. Thus, we presume that the observations at $35.0~\rm GHz$
provide us with a more reliable flux scale. The bandpass was determined from
3C~286 observations on May 12 where we used their spectral window
`{\small SPW}=0' to calibrate our `{\small SPW}=0' and vice versa with `{\small
SPW}=1'. This provided the best possible solution for the bandpass as checks
of calibrated phases and amplitudes of our phase calibrator indicated.
Of course, we are aware of the limitations that our flux density scale
approximation has, but as we will see below, the RC spectra we
obtain show that our flux density measurements in $Ka$-band are broadly
consistent with those we obtained from $K$-band, which were unaffected by the
missing 3C~286 observations. We will discuss this point in Section~\ref{subsec:int}
in more detail, where we will investigate the spectral behaviour of our RC data.
\subsection{VLA archive observations}
\label{subsec:vla_archive}
Observations with the VLA prior to the correlator upgrade were taken in 2006
in $L$- ($1.4~\rm GHz$) and $C$-band ($4.9~\rm GHz$) and published by
\citet{croston_07a}. The $L$-band data were taken in A- and B-configuration,
so that the resolution was matched with the $C$-band data taken in B- and
C-configuration. For both bands the observations were taken in continuum mode
with two intermediate frequencies (IFs) with 50~MHz bandwidth each. The data
were reduced in {\small AIPS} with self-calibration in phase and amplitude (for
$L$-band) and phase calibration in $C$-band. The L-band data were combined
with archive data (AB923) for better (u,v)-coverage. Details of the data
reduction are given by \citet{croston_07a}. We also created a high resolution
map in $C$-band by combining so far un-published A-configuration data with
those in B- and C-configuration.
We used {\small CASA's} MS--MFS clean algorithm \citep{rau_11a} to create new images
using the original reduced (u,v)-data. We used no fitting of the spectral
index as our relative bandwidth is small and there are not enough channels to
do so. However, the multi-scale option results in much smoother images,
particularly of the $C$-band data, in comparison with the {\small AIPS} cleaned
version using only one scale. We used a Briggs robust weighting ({\small
ROBUST}=0). The r.m.s.\ noise of the $L$-band image is $28~\mu\rm Jy\,beam^{-1}$ and
$27~\mu\rm Jy\,beam^{-1}$ for $C$-band. The theoretical expected r.m.s.\ noise is
$25~\mu\rm Jy\,beam^{-1}$ both for $L$- and $C$-band. Hence, our measured
noise level is only $10~\rm per~cent$ higher than the thermal noise level.
We created a set of images with a matched (u,v)-coverage to be sensitive to
the same angular scale of emission. For this we used a {\small UVRANGE} of 0--$76~\rm k\lambda$, which matches the $K$-band ($22~\rm GHz$)
observations, which have the lowest angular resolution. For $L$-band we also
used robust weighing more inclined to natural weighting ({\small ROBUST}=0.5)
to have an angular resolution of $2.5~{\rm arcsec}$, close to the $K$-band
observations. The same was done for the $Ka$-band observations. The shortest
baseline in D-configuration is 35~m (ignoring projection effects at low
elevations), so that the largest angular scale (LAS) is $44~\rm arcsec$ in
$Ka$-band. NGC~3801 is dominated by two radio lobes which have a maximum
angular extent of $\approx 30~\rm arcsec$. Hence, we expect to be able to image
all extended emission at $Ka$-band. At all other observing bands the LAS
measurable by the instrument is significantly larger.
\section{Morphology of the radio continuum emission}
\label{sec:morphology}
\begin{figure*}
\includegraphics[width=1.0\hsize]{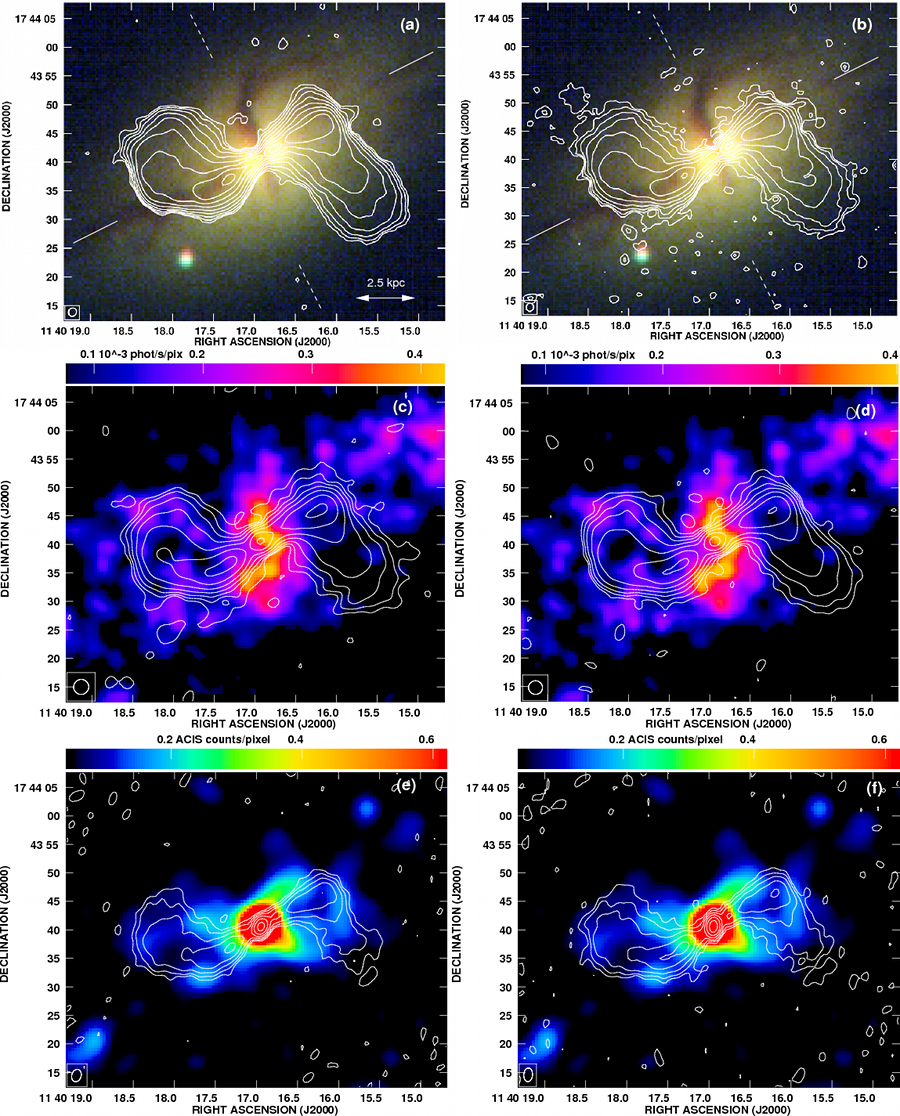}
\caption{RC emission at $1.4$ (a), $4.8$ (b), $22.0$
  (c), $25.0$ (d), $32.3$ (e) and $35.0~\rm GHz$ (f). Contours
  are at (3, 6, 12, 25, 50, 100, 200, 400, 800) $\times$ the r.m.s.\ noise
  level (for some maps the highest contours are not shown). The size of the
  synthesized beam is shown in the lower left corner. The contours are
  overlaid on to false-colour images, which are based on a three-colour
  composite of SDSS $r'$-, $g'$- and $u'$-filter data (a--b), \emph{GALEX} FUV
  data ($\rm FWHM= 5~\rm arcsec$) at $\lambda$155~nm in units of $\rm
  10^{-3}~photons\,s^{-1}\,pixel^{-1}$ (c--d) and \emph{Chandra} X-ray data
  ($\rm FWHM=1.97~{\rm arcsec}$) between 0.5--$5~\rm keV$ in units of ACIS
  counts $\rm pixel^{-1}$ (e--f). The pixel size is $0.5\times 0.5~{\rm
    arcsec^2}$. In panel (a) and (b) we have indicated the approximate major
    and minor axis by solid and dashed lines, respectively.}
\label{fig:high-res}
\end{figure*}

In Fig.~\ref{fig:high-res} we present the distribution of the total power RC
emission for our six observing frequencies. The angular
resolution\footnote{Throughout this paper we refer to the angular resolution
  of a map as the full-width-half-mean (FWHM) of the synthesized beam (RC) or
  the point-spread-function (X-ray and FUV).}, r.m.s.\ noise level, and
further details of the observations are tabulated in
Table~\ref{tab:VLA}. Figures~\ref{fig:high-res}a and b show an overlay on a
three-colour composite made from SDSS $r'$- (red) $g'$- (green) and
$u'$-filter (blue) data. Figures~\ref{fig:high-res}c and d show an overlay on
\emph{GALEX} FUV data at $\lambda$155~nm, which we convolved with a Gaussian
kernel to a resolution of $5~\rm arcsec$. For the false-colour representation
of the FUV data we used a lower cut-off of $5\times$ the r.m.s.\ noise
level. The data used for the overlays in Figs.~\ref{fig:high-res}e and f are
\emph{Chandra} X-ray data between $0.5$--$5~\rm keV$, which were convolved
with a Gaussian kernel to $1.97~{\rm arcsec}$ resolution \citep[see][for
further details]{croston_07a}. The radio emission is clearly contained within
the lobes and the wings, which show very sharp edges. We do not see any more
extended emission at $1.4~\rm GHz$, where the CRe lifetime is the longest. At
this low frequency the emission is almost as well contained as at $4.9$~GHz
and similarly at $22.0$ and $25.0$~GHz. Only at the highest frequencies,
$32.3$ and $35.0$~GHz, the RC emission is contained in a smaller area than at
the lower frequencies. This can be explained by the lower signal-to-noise
ratio of the high frequency observations, which have a similar r.m.s.\ noise
level but a reduced peak flux density than the other maps.
The nucleus clearly has a different spectral behaviour from the radio
lobes. It is not visible at $1.4$ and $4.9~\rm GHz$ against the bright
background emission from the jet, but at the higher frequencies it becomes
ever more prominent and at $32.3$ and $35.0~\rm GHz$ it is the peak of the
RC emission. On both sides, the jet emanates from the nucleus and
then blends into the radio lobes. We notice that the maximum of the RC
emission in the lobes is not coming from an area close to the nucleus but from
an area further away, about 5--10~arcsec (corresponding to $1.3$--$2.5~\rm kpc$)
away from the nucleus. Both lobes have extensions, known as wings, where the eastern wing extends at an angle of $\approx
45\degr$ to the north-east and the western wing extends at angle of $\approx
90\degr$ to the south. Where the jet termination points are located, assuming
a straight jet from the nucleus to the end of the lobes, the contour lines
are compressed (this is best visible in the $1.4$~GHz map) indicating some
interaction with the surrounding ISM. There is a faint
extension at the termination point of the western lobe, both in the $1.4$
and $4.9$~GHz map, but no such extension is seen at the termination point of
the eastern lobe -- particularly in the $1.4$~GHz map, although there is some
spurious emission at $4.9$ and $22.3$~GHz where the structure is less well
defined and does not resemble an extension by the jet.
The positions where the jet has its termination point and bends do correspond
to the shells that are seen in X-ray emission, as can be seen from the overlays
of the RC emission on to the \emph{Chandra} X-ray emission
(Figs.~\ref{fig:high-res}e and f). This suggests a strong connection between
the cosmic-ray population in the jets responsible for the RC emission and the
hot X-ray emitting gas, as discussed by \citet{croston_07a}. We will return to
this point in Section~\ref{dis:age}.

\section{Radio spectral analysis}
\subsection{Radio lobes}
\label{subsec:int}
\begin{figure}
  \includegraphics[width=1.0\hsize]{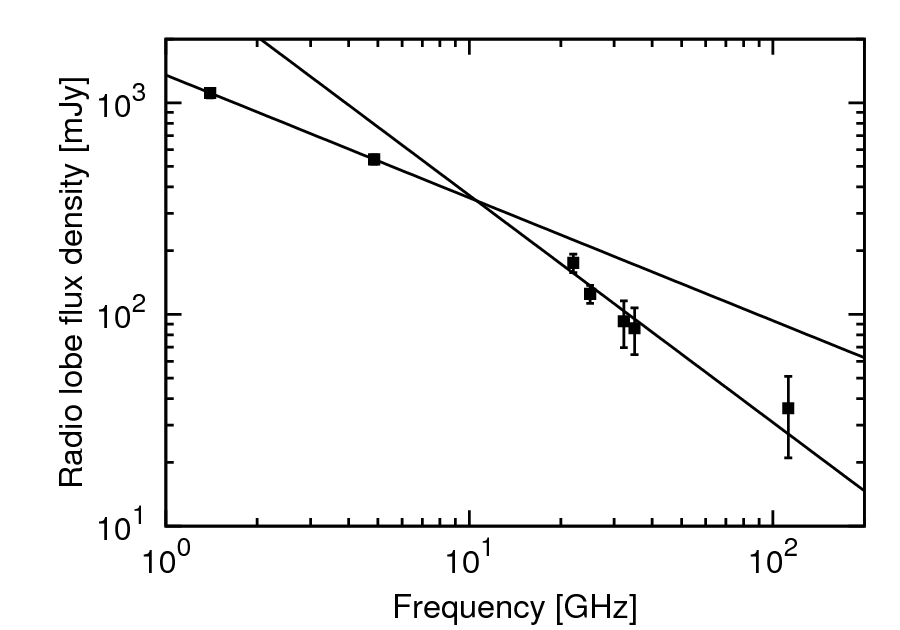}
\caption{Integrated flux density of both radio lobes together as a function of
  frequency. Solid lines show a least-square fit to the data as described in the text.}
\label{fig:flux_int}
\end{figure}

For the analysis of the radio spectral index we convolved our RC maps to a
matched resolution with a Gaussian kernel ($\rm FWHM=2.66~{\rm arcsec}$). The
flux densities were integrated within a rectangular area, using {\small
  IMSTAT} (part of {\small AIPS}). The flux density of the nucleus was
determined by fitting a Gaussian function to it using {\small IMFIT} (part of
{\small AIPS}). The values for the integrated flux density, for the nucleus
and for the lobes are tabulated in Table~\ref{tab:flux}. As the nucleus has a
very different spectral behaviour, we subtract the flux density of the nucleus
to measure the integrated flux density of the lobes only. We combine our data
with those of \citet{das_05a} who used the BIMA interferometer to measure flux
densities in the sub-millimetre wavelength regime. We note that we have
adjusted their quoted flux density error estimate of 10--20 per cent to take
account of the fact that the subtraction of the nucleus in their lower
resolution maps ($\rm FWHM\approx 10~\rm arcsec$) is difficult, introducing
further uncertainties. To separate the continuum emission from the radio lobes
and the nucleus, they scaled a $1.4~\rm GHz$ map with matching resolution to
$112.4~\rm GHz$, assuming a constant spectral index of $0.77$. However, the
spectral index varies across the lobes with an amplitude that we infer to be
$\pm 0.05$ \citep[][their fig.~2]{das_05a}, which introduces a further error
of $8~\rm mJy$. Assuming a $20~\rm per~cent$ calibration error the flux
density of the radio lobes is $(36\pm 15)~\rm mJy$. For the flux density of
the nucleus, the subtraction is not so critical since the emission is not
distributed, so that we only assume a calibration error of $20~\rm per~cent$.
\begin{table}
\centering
\caption{Flux densities as function of frequency.\label{tab:flux}}
\begin{tabular}{rrrr}
\hline\hline
$\nu$ &$S_{\rm tot}$ & $S_{\rm nucleus}$ & $S_{\rm lobes}$  \\ 
$[\rm GHz]$ & $[\rm mJy]$ & $[\rm mJy]$ & $[\rm mJy]$\\
\hline
$1.4$ & $1110 \pm 60$ & $< 6.0\pm 3.0$ & $1110.0 \pm 60$ \\
 $4.9$ & $540 \pm 30$ & $4.5\pm 1.0$ & $535 \pm 30$ \\
 $22.0$ & $183 \pm 18$ & $8.3\pm 1.0$ & $175 \pm 18$ \\
 $25.0$ & $132 \pm 13$ & $6.8\pm 1.0$ & $125 \pm 13$ \\
 $32.3$ & $101 \pm 25$ & $8.5\pm 2.1$ & $93 \pm 23$ \\
 $35.0$ & $94 \pm 24$ & $8.3\pm 2.1$ & $86 \pm 22$ \\
 $86.5$ & $-$ & $15.1\pm 5.0$ & $-$\\
 $110.2$ & $-$ & $10.6\pm 3.3$ & $-$\\
 $112.4$ & $51 \pm 10$ & $15.5\pm 5.2$ & $36.0\pm 15.0$\\
 \hline
\multicolumn{4}{l}{\emph{Notes.} Total integrated flux density, flux density of
the nucleus}\\
\multicolumn{4}{l}{and of the radio lobes. Data
for $\nu \geq 86.5~\rm GHz$ are from}\\
\multicolumn{4}{l}{\citet{das_05a}.}
\end{tabular}
\end{table}
The integrated spectrum of the RC emission in the radio lobes is presented in
Fig.~\ref{fig:flux_int}. It is obvious that the spectrum can not be
described by a single power-law fit, corresponding to a constant spectral
index. Rather, the spectrum shows some significant steepening at frequencies
larger than 10~GHz leading to a curved spectrum. We have therefore fitted two
different power-laws to the data, one between 1 and 10~GHz, and one between 20
and $112.4~\rm GHz$. The resulting spectral indices are $\alpha_{\rm low}=0.6\pm 0.1$
and $\alpha_{\rm high}=1.1 \pm 0.2$. Here, we have defined the radio spectral
index $\alpha$ by $S_\nu \propto \nu^{-\alpha}$, where $S_\nu$ is the radio
continuum flux density at the observing frequency $\nu$. We notice that a
least-square fit to the VLA data alone between 20 and 35~GHz leads to a
much steeper spectral index of $\alpha_{\rm high}=1.6\pm0.4$. Particularly
between $22.0$ and $25.0~\rm GHz$, the spectral index is very
steep. Nevertheless, within the error bars, a single power law can be fitted
through all five data points with $\nu\geq 22.0~\rm GHz$. This means that the flux densities
$Ka$-band at $32.3$ and $35.0~\rm GHz$ are consistent with the other
measurements, lending support to our chosen way of establishing a flux density
scale for the $Ka$-band data (Section~\ref{subsec:jansky_vla}). We hence
proceed with using the $Ka$-band data in our spatially resolved analysis of the spectral
ages, which we present in Section~\ref{subsec:spectral_models}.
\subsection{Nucleus}
\label{subsec:nucleus}
\begin{figure*}
\includegraphics[width=0.8\hsize]{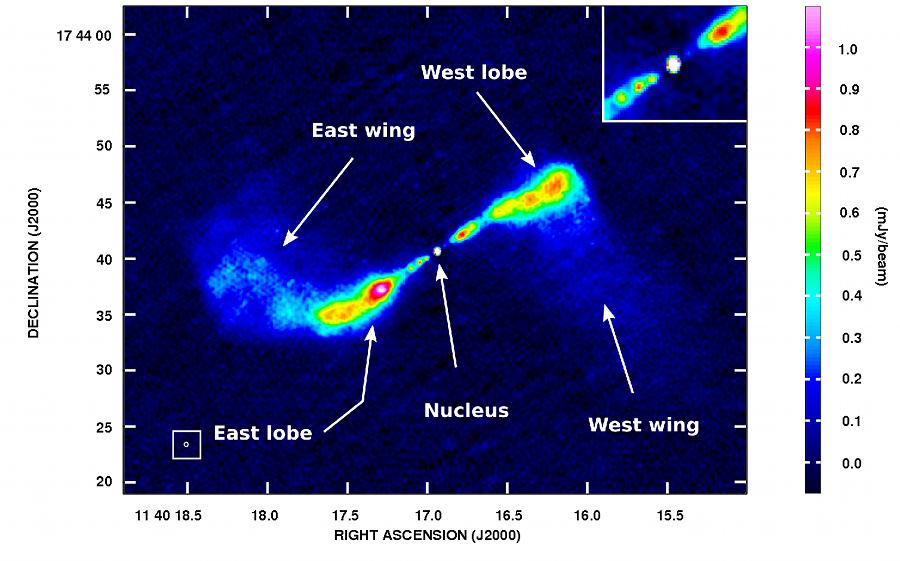}
\caption{NGC~3801 total power RC emission at 4.9~GHz at
  $0.40\times 0.34~{\rm arcsec}$ ($P.A.=2\fdg 9$) angular resolution. The inset shows the
  central part magnified by a factor of two. The size of the synthesized beam
  is indicated by the boxed ellipse in the lower left corner.}
\label{fig:n3801_6cm_ABC-conf}
\end{figure*}

The nucleus has a very different radio spectral index behaviour. It is not
detected at $1.4~\rm GHz$ among the bright background emission of the jet, so
that we can only derive an upper limit for its flux density. At $4.9~\rm GHz$
the nucleus can be detected with difficulty in our map made from $B$- and
$C$-configuration data. However, our high resolution $4.9~\rm GHz$ map made
from a combination with $A$-configuration data, presented in
Fig.~\ref{fig:n3801_6cm_ABC-conf}, shows the nucleus clearly resolved. At
higher frequencies, between $22.0$ and $35.0~\rm GHz$, the nucleus becomes
ever more prominent. In the BIMA map at $112.4$~GHz, presented by
\citet{das_05a}, it has a peak flux density twice as high as the radio
lobes. A Gaussian fit using {\small IMFIT} to the high-resolution $4.9~\rm GHz$
map allows us to determine its position as $\rm R.A.~11^h 40^m 16^s.939$ $\rm
dec.~17\degr 43\arcmin 40\farcs 64$ ($\rm J2000.0$). This position agrees with
that measured by \citet{das_05a} from the BIMA map, at least within the
estimated error which is $1~\rm arcsec$ for the BIMA position. It agrees also
within the quoted position uncertainty of $0.4~{\rm arcsec}$ with the position of the
X-ray nucleus at $\rm R.A.~11^h 40^m 16^s.942$ $\rm dec.~17\degr 43\arcmin
40\farcs 98$ \citep[Chandra Source Catalog,][]{evans_10a}. The optical
position at $\rm R.A.~11^h 40^m 16^s.9$ $\rm dec.~17\degr 43\arcmin 40\farcs
5$ \citep[2MASS Redshift Survey,][]{huchra_12a} agrees within $1~\rm arcsec$ with
the position of the radio nucleus. We can test the pointing error of the VLA
by measuring the position of the nucleus and comparing it with the above
position from the high-resolution map, where we find that between $4.9$ and
35~GHz the positional accuracy is better than $0.2~{\rm arcsec}$.
\begin{figure}
\includegraphics[width=1.0\hsize]{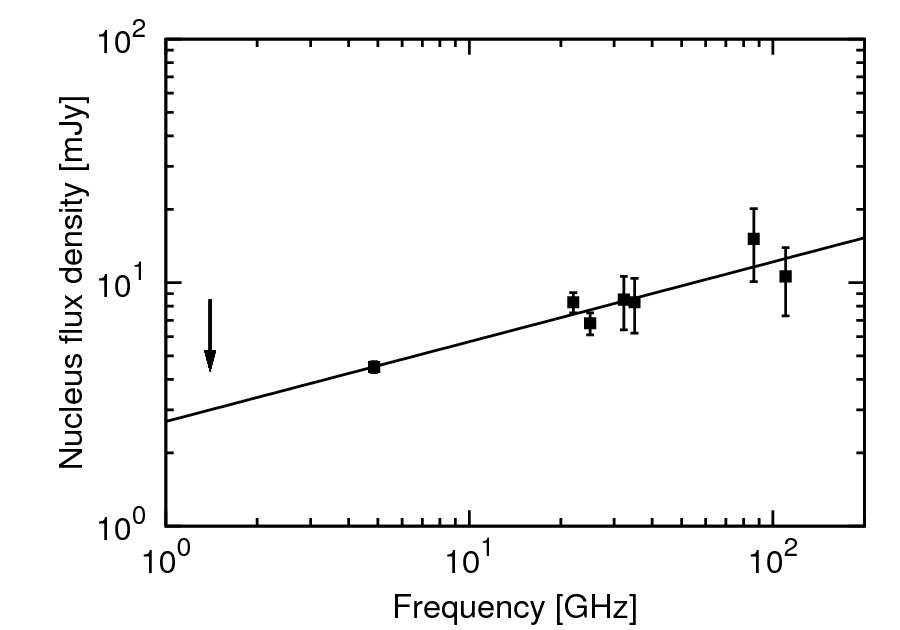}
\caption{Flux density of the nucleus as a function of frequency. The solid line is
  a least-square fit to the data. The arrow indicates the position of the
  upper limit of the flux density at $1.4~\rm GHz$.}
\label{fig:core}
\end{figure}
We present the spectrum of the nucleus in Fig.~\ref{fig:core}, where we
include two more data points from the BIMA observations of \citet{das_05a}. We
can see that the nucleus has clearly an inverted spectrum. A least-square fit
to the data points results in $\alpha_{\rm nuc}=-0.3\pm0.1$. The inverted
spectrum can be explained by an optically thick nucleus either due to
synchrotron or free-free self-absorption. In case of synchrotron
self-absorption we would expect a frequency dependence of the RC intensity as
$I_\nu\propto \nu^{5/2}$ \citep[e.g.,][]{hughes_91a}, and for free-free
absorption we would expect $I_\nu \propto \nu^2$. The fact that both jets have
approximately the same brightness suggests that relativistic beaming towards
the observer plays only a minor role. This means that the jet axis is close to
the plane of the sky and we are looking at viewing angles
$\Theta>>\gamma^{-1}$ on to the jet axis, where $\gamma$ is the Lorentz factor
of the relativistic flow in the jet. Thus, we expect the RC emission near the
nucleus to be partially optically thin synchrotron emission, with spectral
indices of $-0.5\leq \alpha \leq -0.2$ \citep{marscher_80a}. Our observed spectrum
is consistent with this.
\subsection{Spectral index distribution}
\label{subsec:spix}
\begin{figure*}
\includegraphics[width=0.9\hsize]{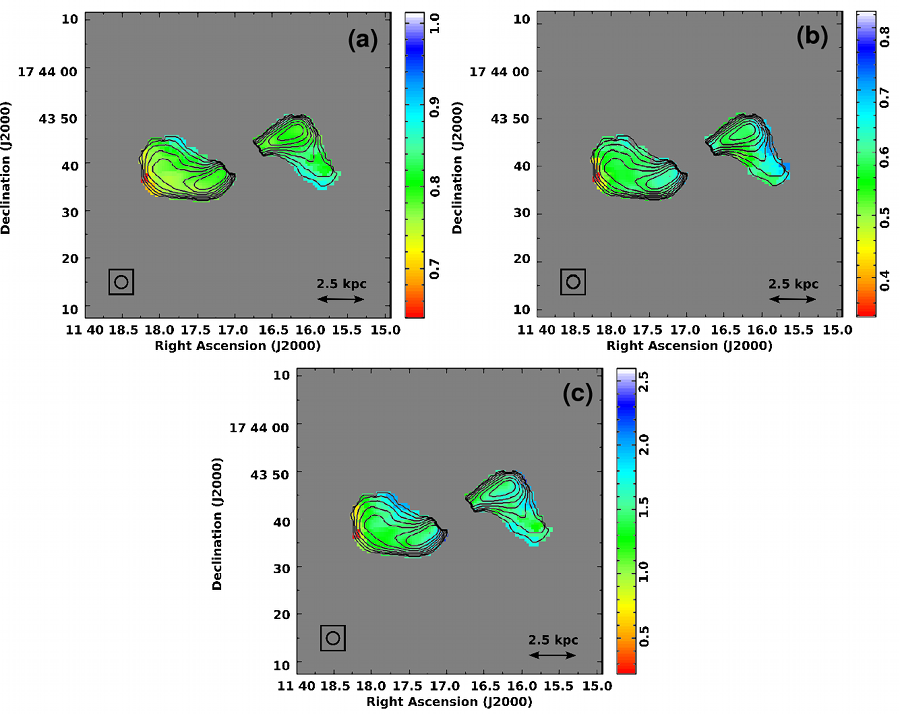}
\caption{Distribution of the radio spectral index. (a) between $1.4$ and
  $35.0~\rm GHz$. (b) between $1.4$ and $4.9~\rm GHz$. (c) between $22.0$ and
  $35.0~\rm GHz$. The radio spectral index $\alpha$ is defined by $I_{\nu}
  \propto \nu^\alpha$, where $I_\nu$ is the RC intensity at the observing
  frequency $\nu$. The angular resolution is $2.66~{\rm arcsec}$, as indicated
  by the circle in the lower left corner. Contours show the RC
  emission at $35.0~\rm GHz$, where the contours are at ($0.57$, $0.74$,
  $1.0$, $1.4$, $1.9$, $2.6$, $3.6$) $\rm mJy\,beam^{-1}$.}
\label{fig:specindex}
\end{figure*}

In Fig.~\ref{fig:specindex} we present the distribution of the radio
spectral index at a angular resolution of $2.66~{\rm arcsec}$ ($\rm = 0.67~kpc$). As
explained in Section~\ref{subsec:vla_archive}, the maps were imaged with a similar
(u,v)-coverage and convolved with a Gaussian kernel to the same resolution, so
that they are sensitive to emission on the same angular scale. The maps were
then registered to the same grid using {\small HGEOM} (part of {\small AIPS}), where the
alignment is accurate within $0.2~{\rm arcsec}$ (Section~\ref{subsec:nucleus}). We used
the program {\small BRATS} \citep[Broadband Radio Analysis ToolS,][]{harwood_13a}
to compute the radio spectral index. This program selects regions within a
given set of maps that have emission above a certain signal-to-noise
cut-off. Where the flux densities are sufficiently high, the region size is
small of the order of one or only a few pixels. In areas of lower flux
densities, the region size can be as large as one beam size to increase the
amount of flux in it and thus to improve the signal-to-noise ratio. A
least-square fit is applied to each map, taking into account the error of the
flux density scale and the variation of emission within a certain region. We
assumed a flux density error of $5~\rm per~cent$ ($\rm \nu\leq 4.9~GHz$), $10~\rm per~cent$ ($\rm 22\leq
\nu\leq 25~GHz$), and $25~\rm per~cent$ ($\rm 32.3\leq \nu\leq 35~GHz$) and a
signal-to-noise cut-off of 10. We notice that the $32.3$ and $35.0~\rm GHz$
maps dominate the signal-to-noise cut-off and we have applied the same regions
to all spectral index maps to ease the comparison between them. We excluded
the nucleus in the spectral index study as it has a very different spectral
behaviour, as discussed in Section~\ref{subsec:nucleus}.
In Fig.~\ref{fig:specindex}a the radio spectral index between $1.4$ and
$35~\rm GHz$ is shown. We can see that the spectral index lies in a narrow range
between $0.65$ and $0.9$ with little variation in most parts of the lobes
where it is around $0.8$. The flattest spectral indices are found near the
termination point of the eastern jet with values between $0.65$ and $0.7$,
whereas the steepest spectral indices are found at the southern tip of the
western lobe with values reaching up to $0.9$.
As we have seen already in Section~\ref{subsec:int}, we can fit the RC spectrum by
two power laws on both sides $\nu = 10~\rm GHz$, where the spectrum
steepens. In Figs.~\ref{fig:specindex}b and c we present the radio spectral
index between $1.4$ and $4.9~\rm GHz$ and between $22.0$ and $35.0~\rm GHz$,
respectively. The spectral index between $1.4$ and $4.9~\rm GHz$ is mostly
ranging between $0.5$ and $0.6$ in the lobes. Close to the eastern termination
point, the spectral index is flatter with $\alpha\approx 0.4$. The western
lobe has a somewhat steeper spectral index at the western edge, where the
spectral index has values between $0.6$ and $0.8$. In contrast, the radio
spectral index between $22.0$ and $35.0~\rm GHz$ is much steeper with values
between $1.2$ and $1.7$. There is again little variation of the spectral index
as function of position within the lobes, with the flattest spectral index
close to the eastern termination point ($0.4\leq \alpha \leq 0.8$). 

Interestingly, at the termination of the western jet the spectral index is
even steeper than in the surrounding area, with values between $2.0$ and
$2.5$. Such a steep spectral index is otherwise only found at the western tip
at the eastern lobe close to the nucleus. \cite{hota_06a} find an east-west
asymmetry in the radio spectral index distribution of the southern radio lobe
in NGC~6764 and attribute this to a possible asymmetry in the distribution of
thermal gas, which contributes a flat RC via free-free emission. The
H~$\alpha$ flux of $8.9\times 10^{-14}~\rm erg\,s^{-1}\,cm^{-2}$, measured
with \emph{HST} by \citet{verdoes_kleijn_99a}, results in a thermal RC flux
density of only 1~mJy at 1~GHz \citep[using the conversion by,
e.g.,][]{deeg_97a}. This is negligible even at 35~GHz, where the lobes have a
flux density of 86~mJy. However, \emph{Spitzer} $8~\mu\rm m$ dust/PAH and
\emph{GALEX} FUV emission show that star formation is present, which is not
seen in the \emph{HST} H~$\alpha$ map. It is hence conceivable that some
ionized hydrogen is present, enshrouded by dust, which would be visible as
thermal RC emission. \citet{croston_08a} found that the asymmetry in the radio
spectral index in NGC~6764 can be seen in the hardness ratio of the X-ray
emission as well. They explained this with shock heating of the gas in the
ISM, which leads to an increase in X-ray temperature and to a flattening of
the RC spectrum due to localized CR re-acceleration. In NGC~3801, however, the
X-ray emitting shell of hot gas surrounding the eastern lobe has a lower
temperature of $0.7~\rm keV$ than the western shell ($1.0~\rm keV$). The
origin of the radio spectral index asymmetry seen near the jet termination
points thus remains unclear.
For the resolved radio spectral index distribution we hence can confirm the
results we found for the integrated emission in Section~\ref{subsec:int}: we find a
flat spectral index of $\alpha_{\rm low}\approx 0.6$ between $1.4$ and
$4.9~\rm GHz$ and a steep spectral index of $\alpha_{\rm high}\approx 1.3$
between 22 and 35~GHz. We find only little spatial variation of the spectral
index across the radio lobes, indicating that the CRe population has an
evolutionary history that varies only little as function of position. In
Section~\ref{subsec:spectral_models} we will use this information to derive a spatially
resolved map of the age of the CRe population within the radio lobes.
\section{Spectral ageing}
\label{sec:ageing}
\subsection{Synchrotron emission}
\begin{figure}
\includegraphics[width=1.0\hsize]{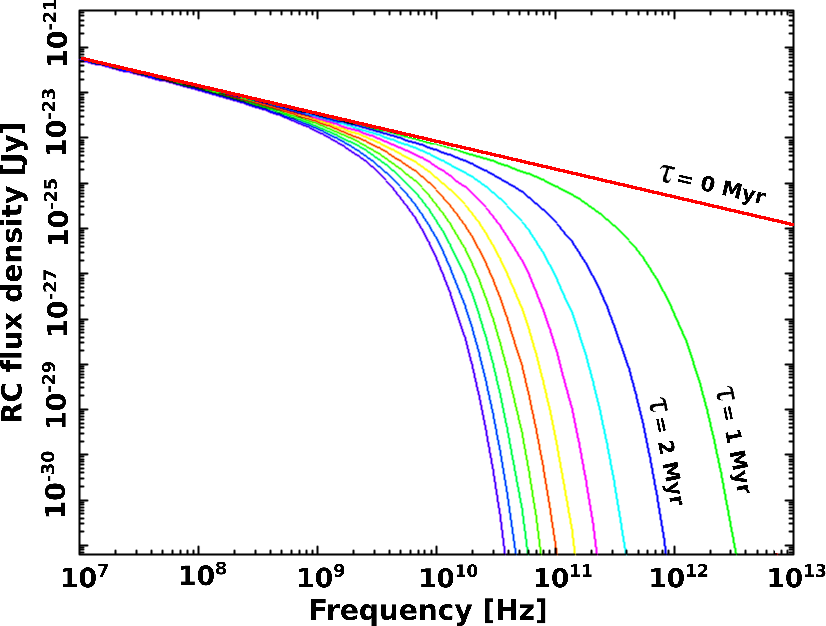}
\caption{Sample JP model spectra for a total magnetic field strength of
  $B=11~\mu\rm G$ between 0 and 10~Myr with an arbitrary
  normalization. The red line at the top corresponds to a CRe age of 0~Myr and
  each line increases by 1~Myr.}
\label{fig:jpmodel}
\end{figure}

Lobes of radio galaxies are sites of particle acceleration, where cosmic rays
can be generated that could be detected directly in Gamma-ray emission
\citep{hardcastle_11a}. Recently, observations were able to find strong
evidence of shock heating in the boundaries of outflowing radio lobes, similar
to particle acceleration by supernova remnants
\citep[SNRs,][]{croston_09a,mingo_11a}. We will now work with the assumption
that we are observing the synchrotron emission of shock accelerated CRe,
similar as in Galactic (or extra-galactic) SNRs. This assumption is supported by
our finding above that the radio spectral index between $1.4$ and $4.9~\rm
GHz$ is with $0.6$, consistent with that what is found for galactic SNRs
\citep{green_09a}. At higher frequencies, the electron show signs of ageing
which is caused by the dominant synchrotron and IC radiation,
which are dependent on the square of the electron energy: the electrons with the
highest energy are losing their energy fastest. This results in a significant
steepening of the radio spectrum above a certain break frequency $\nu_{\rm
  brk}$. The break frequency is as a function of the spectral age $\tau$,
which is the time that has elapsed since freshly
accelerated CRe were injected.
The synchrotron emission of a single CRe has a broad peak at
the critical frequency $\nu_{\rm cri}$, which is related to the electron
energy $E$ and the strength of the magnetic field component perpendicular to
the line-of-sight $B_\perp$ \citep[e.g.,][]{rybicki_86a}:
\begin{equation}
\nu_{\rm cri} = 16.1 \cdot E (\rm GeV)^2 B_\perp(\mu\rm G) ~{\rm MHz}.
\label{eq:nucrit}
\end{equation}
The radiation losses by synchrotron and IC radiation of a
single CRe are given by:
\begin{equation}
\frac{dE}{dt} = -\frac{4}{3} \sigma_{\rm T} c \left (\frac{E}{m_{\rm e} c^2}
\right )^2 (U_{\rm rad} + U_{\rm B}),
\label{eq:enloss}
\end{equation}
where $U_{\rm rad}$ is the radiation energy density, $U_{\rm B} = B^2 / (8\pi)$
is the magnetic field energy density, $\sigma_{\rm T} = 6.65 \times 10^{-25}\,\rm
cm^2$ is the Thomson cross section, and $m_{\rm e} = 511~\rm keV\,c^{-2}$ is
the electron rest mass. Equation~\ref{eq:enloss} is independent of the
electron pitch angle $\theta$, because it assumes that the pitch angles are
isotropic, as expected for self-generated magnetic turbulence induced by
cosmic-ray streaming \citep{kulsrud_69a}. 
The RC spectrum as a function of time can be numerically integrated
using the CRe energy distribution as presented in \citet{kardashev_62a} and
the standard synchrotron radiation formula found in \citet{pachol_70a}. It steepens above the break frequency
$\nu_{\rm brk}$ for which we find \citep[e.g.,][]{hughes_91a}:
\begin{equation}
  \nu_{\rm brk} = 2.52 \times 10^3 \frac{[ B /10~\mu{\rm G}]} {([B / 10~\mu {\rm G}]^2
    +  [B_{\rm CMB} /10~\mu{\rm G}]^2)^2[\tau / \rm Myr]^2}~{\rm GHz} \mbox{.}
\label{eq:nu_brk}
\end{equation}
Here, the equivalent cosmic microwave background (CMB) magnetic field strength
$B_{\rm CMB} = 3.2(1+z)^2~\rm \mu G$ is defined so that the magnetic energy
density is equal to the CMB photon energy density. The break frequency moves
with time to lower frequencies. In Fig.~\ref{fig:jpmodel} we show a
selection of JP model spectra, with spectral ages between 0 and 10~Myr with an
increment of 1~Myr. We notice that the RC spectrum steepens at much lower
frequencies than the actual break frequency, so that model spectra have to be
fitted to the data. For instance, for a spectral age of 2~Myr the break
frequency is $\nu_{\rm brk}=400~\rm GHz$, much higher than the frequency of
10~GHz above where we found a steepening of the RC spectrum
(Section~\ref{subsec:int}). In order to calculate the synchrotron spectrum and the
break frequency we need to estimate the magnetic field strength, which we will
do in Section~\ref{subsec:equip}.
The spectral behaviour above the break frequency is different in various models.
The so-called JP model by \citet{jaffe_73a} assumes that the CRe pitch angles
with respect to the magnetic field are isotropized, whereas the KP
(Kardashev--Pacholczyk) model assumes that the CRe pitch angle is retained from
its injection into the radio lobe. The JP model has the advantage that it
better corresponds to the picture of self-generated turbulence due to
Alfv{\'e}n waves \citep[e.g.,][]{kulsrud_69a}. The CRe pitch angle with
respect to the magnetic field lines will be isotropized in the rest frame of
the magnetized plasma. The break frequency is hardly affected by the choice of
either the KP or JP model. The main difference between the two models is that
the JP model predicts an exponential cut-off, whereas the KP model predicts a
spectral steepening of $\alpha_{\rm high}=(4/3)\alpha_{\rm low}+1$. This is
because there are always some electrons at small pitch angles that still do
emit synchrotron radiation, even for CRe that are observed above the break
frequency. Both the JP and KP model assumes that the CRe are
accelerated and injected in a single burst, which are referred to as
single injection (SI) models. In addition to the single burst injection
models, there are also continuous injection models
\citep[e.g.,][]{kardashev_62a,pachol_70a}. These models show also a spectral
break, but the steepening is with $\Delta\alpha=0.5$ much smaller than what
SI models predict. 
\subsection{Equipartition magnetic field strength}
\label{subsec:equip}
In order to calculate the magnetic field strength the use of the equipartition
assumption between the cosmic ray and the magnetic field energy density is the
standard method \citep[e.g.,][]{beck_05a}. FR~I radio galaxies appear to have
under-pressurized radio lobes if equipartition is used, so that the combined
pressure from cosmic rays and magnetic fields is lower than the surrounding
pressure of the thermal hot gas as measured from X-ray emission \citep
[e.g.,][]{morganti_88a,croston_03a}. In NGC~3801 the shells of hot X-ray
emitting gas, which are upstream of the radio lobes, have a thermal pressure that
exceeds the pressure inside the lobes by a factor of $\approx 2$--5. As the
radio lobes are clearly expanding \citep{croston_07a}, the pressure inside the radio lobes as
determined from equipartition must be an underestimate. In many FR~I radio
lobes the equipartition pressure is even lower than that of the surrounding ISM. However,
NGC~3801 is an exception to this rule as well as Centaurus~A: the pressure in the X-ray
emitting shells and hence the true pressure inside the radio lobes is more
than an order of magnitude higher than in the surrounding ISM
\citep{croston_07a}.  This suggests that the deviation from equipartition may
not be very large and we use the equipartition estimate as a lower estimate
for the magnetic field strength.
We calculated the equipartition minimum internal pressure of the radio lobes,
using measurements of the $1.4~\rm GHz$ flux density for each lobe to
normalize the synchrotron spectrum. Following \citet{croston_07a} we estimate
the volumes of the lobes as spheres, finding for the western lobe $\rm
2.1\times 10^{10}~pc^3$ and for the eastern lobe $\rm 2.5\times
10^{10}~pc^3$. These numbers are uncertain, because it is difficult to
estimate the line-of-sight depth of the radio lobes.  We assumed a broken
power-law CRe number distribution with an injection index of $p=2.0$. This is
the theoretical expectation from first-order Fermi CRe acceleration by strong,
non-relativistic shocks \citep{bell_78a,blandford_78a}. As we will see below
in Section~\ref{subsec:spectral_models}, we can fit for the asymptotic
radio spectral index at low frequencies which is identical to the injection
radio spectral index $\alpha_{\rm inj}$. We find $\alpha_{\rm inj}=0.5$
corresponding to a CRe spectral index of $p=2.0$ ($\alpha_{\rm inj}=(p-1)/2$), in excellent agreement with
the theoretical expectation. We integrated the CRe energy between the Lorentz
factors of $\gamma_{\rm min}=10$ and $\gamma_{\rm max}=10^5$, with a break at
$\gamma_{\rm brk}=1.6\times 10^4$. As usual for radio lobes (both FR~I and
FR~II) we assumed an electron--positron plasma with no contribution from
protons. The resulting magnetic field strength is on average $B=(11\pm
2)~\mu\rm G$, where we assumed a $20~\rm per~cent$ error due to the uncertainty in the
source geometry. In order to have pressure equilibrium with the X-ray shells a
proton dominated model with $K\approx 200$ and $B=(49\pm 10)~\mu\rm G$ is
needed, where $K$ is defined by the ratio of the sum of proton and electron
energy density to that of the electron energy density alone. The
most likely magnetic field strength lies between these two values, which we
take into account when calculating the spectral age in
Section~\ref{subsec:spectral_models}.
The break in the CRe spectrum, and hence $\gamma_{\rm brk}$, depend on the
magnetic field strength, and vice versa. As we have now determined that the
spectrum steepens above 10~GHz (Section~\ref{subsec:int}), we can test whether we
have used the correct $\gamma_{\rm brk}$. From Eq.~\ref{eq:nucrit} we can
estimate the electron energy at $10~\rm GHz$ to $E=7.5~\rm GeV$ (for the model
with no proton contribution). The frequency of 10~GHz thus corresponds to
$\gamma_{\rm brk}= E/(m_{\rm e}c^2) \approx 1.5\times 10^4$, in agreement with
our assumption above.
\begin{figure*}
\includegraphics[width=0.9\hsize]{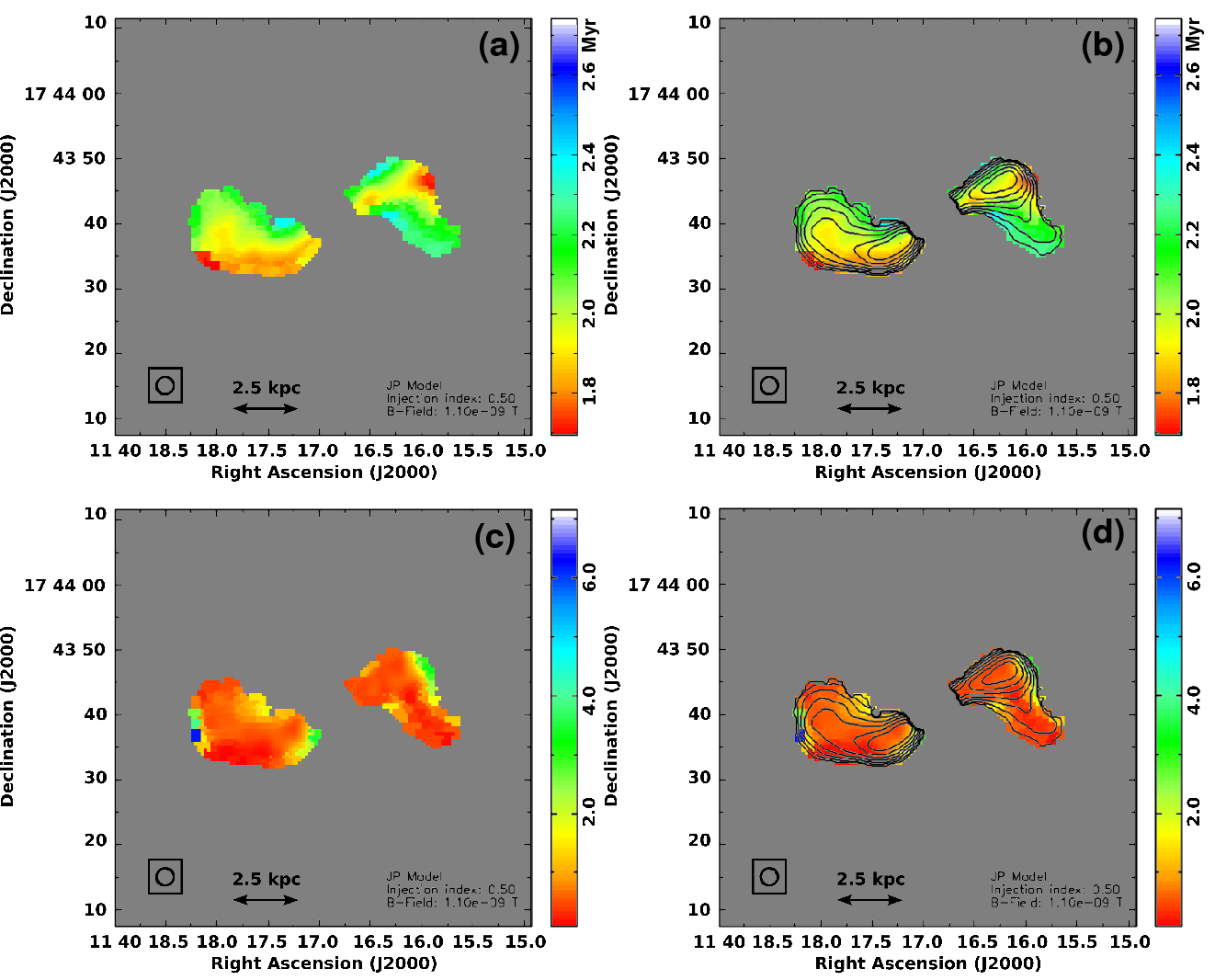}
\caption{CRe spectral age as calculated from a JP model fit to the RC data
  (a+b). The magnetic field strength is $B=11~\mu\rm G$ and the injection
  radio spectral index is $\alpha_{\rm inj}=0.5$. Contours show the total
  power RC emission at $35.0~\rm GHz$, where the contours are at ($0.57$, $0.74$,
  $1.0$, $1.4$, $1.9$, $2.6$, $3.6$) $\rm mJy\,beam^{-1}$. The angular
  resolution is $2.66~{\rm arcsec}$ as indicated by the circle in the lower
  left corner. The bottom panels show the distribution of the reduced $\chi^2$
  (c+d). For details of the fitting procedure see
  Section~\ref{subsec:spectral_models}.}
\label{fig:spec_ageing}
\end{figure*}
\subsection{Spectral models}
\label{subsec:spectral_models}
We used {\small BRATS} to fit JP models to our observed RC maps with the
fitted regions and flux density errors identical to the ones used for
calculating the radio spectral index (Section~\ref{subsec:spix}). A constant
magnetic field strength (throughout this section we use $B=11~\mu\rm G$) is
assumed throughout the source and the injection radio spectral index is
prescribed as well. This leaves as free parameters the absolute flux
  density normalization and the spectral age, which we fit using a
  least-square procedure (with 6 frequencies we have $\rm d.o.f.=4$ as numbers of degrees-of-freedom). We present the
spectral age maps that we have obtained, along with the distribution of the
reduced $\chi^2$ ($=\chi^2/{\rm d.o.f.}$) in Fig.~\ref{fig:spec_ageing}.  We define
  $\chi^2$ by
\begin{equation}
  \chi^2 = \sum \left(\frac{D_i - M_i}{\sigma_i}\right)^2,
\end{equation}
where $D_i$ is the $i$th flux density measurement, $M_i$ the
  corresponding model value, and $\sigma_i$ the error on the measured
  flux
  density. As already found in our
analysis of the radio spectral index in Section~\ref{subsec:spix}, the
variation of the radio spectral index is small across the lobes and so is the
variation of the spectral age. The age lies within a range of $1.8$ and
$2.4~\rm Myr$. The youngest CRe are found near the termination points of both
the western and eastern jet, where the age is only $\approx 1.8~\rm Myr$. The
area between the nucleus and the termination points, where the jet is
presumably located, shows also a slightly younger age $(1.9$--$2.0~\rm Myr)$ than
in most other areas of the lobes $(2.0$--$2.3~\rm Myr)$. Areas where the age is
larger than $2.3~\rm Myr$ are sparse and found only at the outer edge of both
lobes. We notice that these are to be considered as the upper limits of the
spectral age, because we have used the lower limit on the magnetic field
strength from equipartition. For the higher magnetic field strength, the ages
are reduced by a factor of 10. The minimum spectral age, $1.8~\rm Myr$, is
clearly larger than zero. This differs from FR~II radio galaxies that have a
so-called `hotspot' with a spectral age consistent with zero
\citep[e.g.,][]{harwood_13a}. The difference can be explained by the fact that
in NGC~3801 there is no dominating young emission component such as found in
the hotspots of FR~II radio galaxies, because cosmic-ray transport plays an
important role. We will come back to this aspect in more detail in
Section~\ref{dis:age}.

A magnetic field strength of $B=11~\mu\rm G$, with no contribution from
protons, corresponds to a spectral age of 2~Myr in most parts of the radio
lobes, in good agreement with the dynamical age of the lobes, as determined by
\citet{croston_07a} based on the expansion speed of the X-ray emitting
shells. For $B=49~\rm \mu G$, however, the magnetic field strength derived for
a proton dominated model, we find a spectral age of $0.2~\rm Myr$, which can
be regarded as the lower limit. Thus, will use in the following the spectral
ages for the lower magnetic field strength and refer to Section~\ref{dis:age} for a
detailed discussion of the most likely value for the age of the radio lobes.
We do not include the BIMA data point at $112.4~\rm GHz$ when fitting for the
spectral age. We tested the influence of the $112.4~\rm GHz$ data by
extrapolating the radio spectral index between $1.4$ and $35~\rm GHz$ to
$112.4~\rm GHz$. This is possible, because the radio spectral index between
$1.4$ and $35~\rm GHz$ is almost identical to the spectral index between $1.4$
and $112.4~\rm GHz$ \citep{das_05a}. The result is that the spectral age is
reduced by the inclusion of the data point to $1.3~\rm Myr$. This is because
the break frequency is shifted to higher frequencies as the flux density at
$112.4~\rm GHz$ is too high for the JP model fitted to the data points between
$1.4$ and $35~\rm GHz$ only. As explained above, the RC emission
at each location is not caused by a single-age plasma, but by a superposition
of various CRe populations. In this case, we expect to observe a superposition
of JP models, resulting in a more power-law type behaviour for higher
frequencies in agreement with the BIMA observations. Alternatively, a different spectral model
such as the KP model, which does not have an exponential decrease of the flux
density at higher frequencies, may be fitted. A variety of the JP model is the
so-called Tribble JP model, that drops the assumption of a constant magnetic
field within the source \citep{tribble_93a}. \citet{hardcastle_13a} modelled
Tribble JP spectra, assuming a Maxwell--Boltzmann distribution of magnetic field
strengths, and found that the shape of the spectrum at higher frequencies beyond
the break frequency is not as steep as for a JP model. We hence fitted a
Tribble JP model to the data and found that the fit is almost identical to the
pure JP model (reduced $\chi^2=1.1$, $\tau=1.3~\rm Myr$). The same is true for the
physically less motivated KP model. This may come as no big surprise, because
we are fitting only for frequencies much smaller than the break frequency at
$\nu_{\rm brk}\approx 400~\rm GHz$ and the difference between the JP, KP and
Tribble JP model are only pronounced above the break frequency. The same
result was found by \citet{harwood_13a}, who fitted the various
varieties of spectral models to data of two FR~II sources and found only a
small difference between them. The most promising aspect of the Tribble JP
 model is that it can explain the better fitting to the data points, as
for a KP model, but without making the unphysical assumption of preserved
electron pitch angles.
\begin{figure}
\includegraphics[width=1.0\hsize]{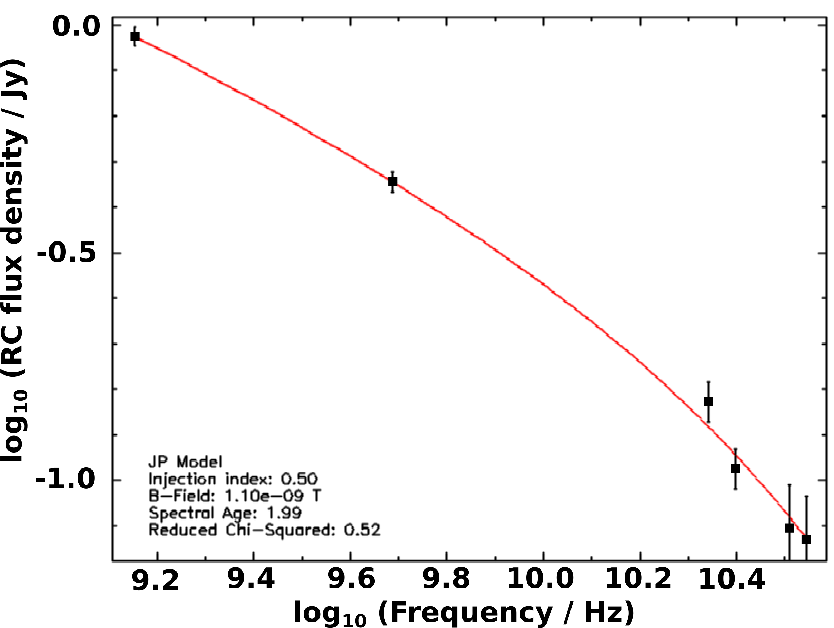}
\caption{Plot of the integrated flux density of the radio lobes as function of
  frequency. Overlaid is the JP model fit to the data used to determine the
  spectral age of the CRe.}
\label{fig:plotmodelobs}
\end{figure}
\begin{figure*}
\includegraphics[width=0.9\hsize]{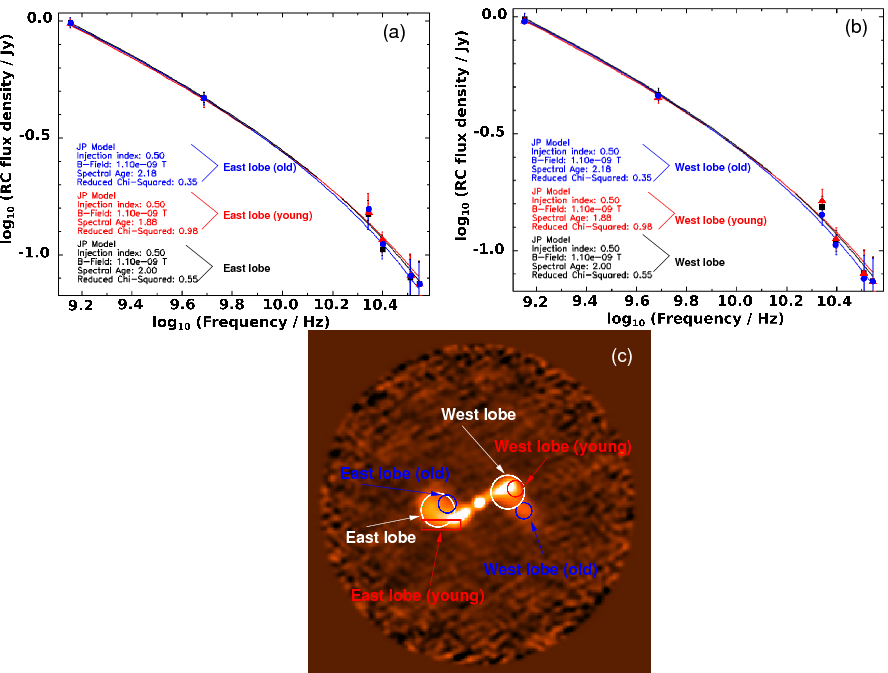}
\caption{Plot of the integrated flux density of the radio lobes as function of
  frequency in various regions within the radio lobes. Overlaid are the JP
  model fit to the data used to determine the spectral age of the CRe. Spectra in the eastern (a) and western (b) radio lobe, respectively.  The
    flux density scale is arbitrary and the spectra have been aligned
    vertically at low frequencies in order to highlight the differences at
    higher frequencies. (c) position of the regions within both radio lobes,
    overlaid on to the $35.0~\rm GHz$ RC emission at $2\farcs 66$ resolution.}
\label{fig:region_spectra}
\end{figure*}
The quality of the fit is mostly very good, with almost all parts of the lobes
having a reduced $\chi^2< 2.0$. Only at the edges of the lobes, where
the signal-to-noise ratio is lowest, is the reduced $\chi^2$ larger with
$2.0<\chi^2< 5.0$. The eastern edge of the eastern lobes has the
highest reduced $\chi^2\approx 8.0$, which is at the area where no spectral
steepening for frequencies larger than 10~GHz is observed
(Section~\ref{subsec:spix}).  We notice that the areas with the youngest spectral
ages, hereafter called the `termination points', do not coincide with a
particular high value of reduced $\chi^2$, nor does the younger area near the
southern edge of the eastern lobe. Only the western edge of the western lobe
has a slightly increased reduced $\chi^2$. The oldest spectral ages are all
located in areas of a low reduced $\chi^2$. We conclude that the variation in
CRe age is not due to a variation in the quality of the fit, supporting the
idea that we are tracing an actual change of the spectral behaviour. We also
varied the injection radio spectral index $\alpha_{\rm inj}$ and found that
$\alpha_{\rm inj} = 0.50$, which is the theoretically expected value,
minimizes the average reduced $\chi^2$ of the regions in our spectral age
maps.
We can also
repeat the spectral age analysis for the integrated flux densities as
measured from the radio lobes. This is a valid procedure because of the small
variations in the point-to-point fitted ages. The resulting spectrum and the JP model fit to
the data are shown in Fig.~\ref{fig:plotmodelobs}. For the integrated
measurements we derive a CRe age of $\tau_{\rm int}=(2.0\pm 0.4)~\rm Myr$ with a
reduced $\chi^2=0.53$. This age is consistent with the spatially resolved
analysis, which as described above shows only little variation of the spectral age across the
lobes with values mostly between $1.9-2.2~\rm Myr$. In
Fig.~\ref{fig:region_spectra} we show spectra of RC emission integrated in
small regions and fitted with a JP model. We plotted the spectra in each lobe
on top of each other to highlight the small changes that can be seen for
frequencies $\nu>\nu_{\rm brk}$, where younger areas show a smaller degree of
spectral steepening as expected. In order to estimate the error on our
electron spectral ages, we varied the input parameters and repeated the fit of
the JP model to the data. This was done for both the resolved and the
integrated spectral age measurements. A change of the injection spectral index
by $\pm 0.05$ has a $20~\rm per~cent$ effect on the derived electron ages. Hence, we
adopt this number as the formal error of the spectral ages.
\section{Discussion}
\label{sec:discussion}
\subsection{Energetics of the radio lobes}
\label{dis:energy}
NGC~3801 is an example of a radio galaxy with kpc-sized lobes that are
surrounded by hot X-ray emitting gas. A key question is whether the radio
lobes are heating the surrounding ISM of its elliptical host galaxy and are
thus transferring the kinetic energy of the black hole powered jets to the
environment. This can be done via shock heating where the radio lobes expand
supersonically into the surrounding gas and are thus heating the gas via
adiabatic compression. Such a model was proposed by \citet{croston_07a} who
found that the density contrast of the X-ray emitting gas across the shock is
a factor of $\approx 4$, in agreement what is expected from Rankine--Hugoniot
shock jump conditions. They found Mach numbers of 3--8, confirming a
supersonic expansion of the radio lobes. The sound speed in the ISM is
$\approx 210~\rm km\,s^{-1}$, so that a Mach number of $M=4$ corresponds to a
lobe expansion speed of ${\rm v}\approx 850~\rm km\,s^{-1}$. If we take the distance
between the nucleus and the re-acceleration site as $3.5~\rm kpc$, using the
distance to the brightest parts of the symmetric radio lobes without the
wings, we can derive the lobe dynamical age as $\tau_{\rm dyn} = L / {\rm v} =
4~\rm Myr$. Our derived spectral age of $\tau_{\rm int}=2.0~\rm Myr$
($B=11~\mu\rm G$) is a factor of two lower than this dynamical age, which can
be considered as a good agreement given the uncertainties involved.
We now discuss the various effects to be taken into account that can alter the
spectral age in comparison to the dynamical age of the lobes. These are
IC and adiabatic losses, cosmic-ray transport (either by diffusion or
convectively in the backflow from the
termination points) and \emph{in situ} re-acceleration. Any additional losses such as
adiabatic losses and IC radiation would mean that the spectral
age is an overestimate of the true lobe age. The ratio of IC
losses to synchrotron radiation is proportional to the ratio of the
radiation energy density $U_{\rm rad}$ to the magnetic field energy density
$U_{\rm B}$. The magnetic field energy density $U_{\rm B}=B^2/(8\pi)$ for
$B=11~\rm\mu G$ (Section~\ref{subsec:equip}) is $U_{\rm B}=4.8\times
10^{-12}~\rm erg\,cm^{-3}$. The radiation energy density consists of a
contribution from the stellar radiation background field and heated dust as
well as the CMB. We
calculated the dust component from the far-infrared luminosity as measured by
IRAS and used an area of $A=\pi (5.6~\rm kpc)^2$ to calculate the dust
component as $U_{\rm dust}=L_{\rm FIR} / (2 A c)$. We scaled the stellar
component to the dust component by $U_{\rm ISRF} = 1.73 \times U_{\rm dust}$ and added
the CMB radiation energy density of $4.2\times 10^{13}~\rm erg\,cm^{-3}$. The
resulting radiation energy density is $U_{\rm rad}=1.1\times 10^{-12}~\rm
erg\,cm^{-3}$, only $20~\rm per~cent$ of the magnetic field energy density. We hence
conclude that energy losses of the CRe due to IC
radiation are small in relation to losses due to synchrotron radiation.
The effect of adiabatic losses on the CRe can be estimated by comparing the
$P{\rm d}V$ work of the bubbles to carve the cavities into the ISM with the
energy content of the radio lobes. \citet{croston_07a} estimate that the total
energy in the shells, the sum of thermal and kinetic energy, is $(1.4\pm 0.4)
\times 10^{56}~\rm erg$. This is in good agreement with the pressure work
available from the internal pressure, which is $P_{\rm int}V\approx
10^{56}~\rm erg$. The energy required to inflate both cavities is only
$\approx 7 \times 10^{54}~\rm erg$, a factor of $\approx 25$ lower than the
energy stored in the shells. This discrepancy can be understood by the large
ratio $P_{\rm int}/P_{\rm ISM}=15$--$23$ of the internal pressure within the
lobes to the pressure of the ISM, assuming that the internal
pressure is equal to the pressure of the thermal X-ray emitting gas in the
shells surrounding the radio lobes. Thus, adiabatic losses are only a small
contribution to the CRe losses and will be neglected in the
following discussion.
Cosmic-ray transport can play an important role in shaping the spectral age
distribution: cosmic rays may diffuse along magnetic field lines or be
transported convectively together with the magnetized plasma.  In order to
determine the structure of the magnetic field within the lobes, linear
polarization measurements would be required; however a study by
\citet{hardcastle_12a} of the FR~I source 3C~305 suggests that there can be
significant ordering of magnetic fields within radio lobes. This would allow
CRe to stream along magnetic field lines with a maximum speed equal to the
Alfv{\' e}n velocity ${\rm v}_{ \rm A} = B/\sqrt{4\pi \rho}$, where $\rho$ is
the gas density. The gas density within the lobes is smaller than that of the
ISM, $n_{\rm ISM}=4.6\times 10^{-3}~\rm cm^{-3}$, so that the Alfv{\' e}n
velocity within the lobes is larger than $560~\rm km\,s^{-1}$, comparable to
the lobe expansion speed. During the dynamic age of the lobes $\tau_{\rm
  dyn}\approx 4 \times 10^6~\rm yr$ the cosmic-ray diffusion length may be as
high as 7~kpc, comparable to the size of the lobes. We hence expect cosmic-ray
diffusion to be able in principle to even out any difference in the spatial
distribution of the CRe spectral age. As an alternative to diffusion, cosmic
rays can also be transported convectively together with the magnetized
plasma. Hydrodynamical simulations of FR~I galaxies by \citet{perucho_07a}
show that the backflow surrounding the radio lobes can have mildly
relativistic velocities with $0.1$--$0.2c$, much faster than the lobe expansion
speed. A convective transport in the backflow is hence a second possibility
to explain the very uniform distribution of CRe spectral ages within the lobes.
If IC and adiabatic losses were significant, we would overestimate the
spectral age because we underestimate the CRe energy losses. The largest
uncertainty comes however from the estimate of the magnetic field
strength. For the higher magnetic field strength of $B=49~\rm\mu G$, which
assumes a pressure contribution from protons, the spectral ages are
approximately a factor of $(49/11)^{3/2}\approx 10$ (see Eq.~\ref{eq:nu_brk})
lower than for the equipartition magnetic field of $B=11~\rm\mu G$ (without
the contribution from protons).
\subsection{Dynamical age of the lobes}
\label{dis:age}
\begin{figure*}
\includegraphics[width=1.0\hsize]{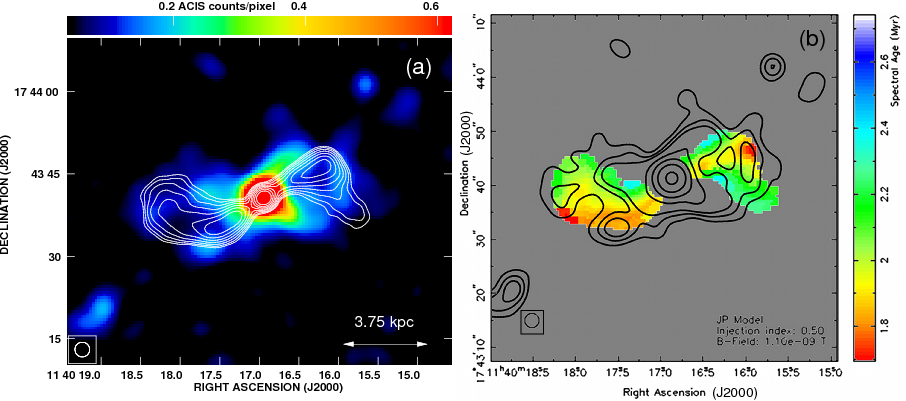}
\caption{(a) Gaussian smoothed ($\rm FWHM=1.97~{\rm arcsec}$) $0.5-5~\rm keV$
  image of the \emph{Chandra} X-ray data, with contours of the RC emission at
  $35.0$~GHz ($\rm FWHM=2.66~{\rm arcsec}$) overlaid. The contours are identical
  to those in Figs.~\ref{fig:spec_ageing}b and d to ease comparison with the
  maps of the spectral age distribution. The units are in ACIS $\rm
  counts\,pixel^{-1}$ with a pixel size of $0.5\times 0.5~{\rm arcsec^2}$. (b)
  Spectral age distribution with X-ray contours of the data shown in (a)
  overlaid. Contours are at ($0.074$, $0.11$, $0.15$, $0.60$, $1.2$, $2.4$)
  ACIS counts $\rm pixel^{-1}$.}
\label{fig:n3801_age_X-ray_fig}
\end{figure*}

Now that we have established that IC and adiabatic losses of the CRe do not
influence our estimate of the spectral age, but cosmic-ray transport does, we
try to constrain the dynamical age of the lobes. In
Section~\ref{subsec:spectral_models} we found that the minimum spectral age is
$1.8~\rm Myr$ ($B=11~\mu\rm G$) and hence larger than zero. This can be
explained by cosmic-ray transport: the CRe population at any given location is
not a single-age plasma but a superposition of CRe populations that have been
transported from various locations within the lobes. This scenario is
corroborated by the fact that the spectral age distribution is almost uniform
across the lobes with only small variations. Moreover, we observe always a
line-of-sight average of the spectral age, further reducing the age
differences within the lobes. Ongoing CRe acceleration, for instance in the
termination points, lowers the spectral age but since the older components are
dominating, the observed variation in the spectral age is small. We refrain
from using a continuous injection model to explain the spectral behaviour of
our source, because such a model is only valid if one emission component
dominates, which is for instance the case in a `hotspot' of a FR~II galaxy
\citep{heavens_87a}. The oldest CRe we find in the radio lobes have an age of
$2.4~\rm Myr$. Thus, the most likely dynamical age lies between the ages of
$2.4$ and $4~\rm Myr$, where the latter value assumes a constant expansion
speed of $850~\rm km\,s^{-1}$. The good agreement between spectral and
dynamical age supports the assumption of an equipartition magnetic field
strength with $B=11~\mu\rm G$. A higher magnetic field strength, where the
magnetic pressure is in equilibrium with the pressure of the X-ray emitting
shells, is unlikely, because the resulting spectral ages would be a factor of
10--20 too low in comparison with the dynamical age.
In Fig.~\ref{fig:n3801_age_X-ray_fig} we present an overlay of the spectral
age map on to the \emph{Chandra} X-ray emission, in order to study the
interplay between the CRe acceleration and the shock heating of the
surrounding gas. We find that the termination points of the spectral age maps
(see Fig.~\ref{fig:spec_ageing}) are indeed spatially close to the X-ray
shells, although the prominent eastern termination point is located within a
gap of the X-ray emission. This strengthens our conclusion, and that of
\citet{croston_07a}, that the X-ray emitting shells are indeed stemming from
shock heating due to the expanding radio lobes. The shocks upstream of the
contact discontinuity heat the surrounding gas of the ISM and the shocks
downstream of the discontinuity re-accelerate the CRe.  We do not have the
spatial resolution to resolve the two regions, but \citet{croston_09a} did
find that the X-ray emission in Centaurus~A is indeed upstream of the RC
emission, displaced by a distance of $\approx 300~\rm pc$.\footnote{In their
  case the X-ray emission is of non-thermal origin by highly energetic CRe
  that are emitting synchrotron radiation in the X-ray wavelength regime. This
  can be understood by the much larger lobe expansion speed of Centaurus~A
  ($2600~\rm km\,s^{-1}$) in comparison to NGC~3801, which allows effective
  particle acceleration in shocks.}
The age variation of the CRe in the lobes allows us to calculate the CRe bulk
speed. The transport length is equal to the length of the wing, which is
$L\approx 2.5~\rm kpc$ for both wings. The age difference between the CRe age
at the termination points, $1.8~\rm Myr$, and the oldest CRe, $2.4~\rm Myr$,
is $\Delta\tau =0.6~\rm Myr$, so that we find ${\rm v}_{\rm bulk}=4200~\rm
km\,s^{-1}$, far higher than the Alfv{\' e}n speed of $560~\rm
km\,s^{-1}$. The Alfv{\' e}n speed is only a lower limit though, because we
have used the gas density of the ISM to calculate it
(Section~\ref{dis:energy}) and the density within the radio lobes is likely
much lower. We can hence not rule out the streaming of cosmic rays along the
magnetic field lines as an alternative mode of transport. However, this would
require a spatial ordering of the magnetic field structure. Linear RC
polarization measurements find a fractional polarization that is generally
less than the theoretical maximum expected for a purely ordered magnetic
field. In the presence of a turbulent component, the transport can be
characterized rather by diffusion, where the required diffusion coefficient of
$D=L^2/\Delta\tau$ is with $D= 3.1 \times 10^{30}~\rm cm^2\,s^{-1}$ a factor
of 10--100 higher than what is observed in the Milky Way or in other galaxies
\citep[e.g,][]{strong_07a,heesen_09a,berkhuijsen_13a,buffie_13a,tabatabaei_13a}. The
more likely scenario is thus that the CRe are indeed transported not by
diffusion but convectively together with the \emph {backflow} of magnetized
plasma. Such a backflow has been proposed to explain the spectral index
gradient in giant FR~II radio galaxies before \citep{alexander_87a}, but is
also seen in hydrodynamical simulations of FR~I radio galaxies similar to
NGC~3801 \citep{perucho_07a}.
Apart from the termination points, we see another area of younger CRe along
the area where the jet probably lies within the lobes.  This could attributed
to re-acceleration either within the jet (i.e.\ close to the nucleus) or
within the radio lobes. Shock waves in the jet can be formed by standing
re-collimation shocks, which are seen in the simulations of
\citet{perucho_07a}. Such shocks may be visible as knots in the RC emission,
particularly at high resolution. In the inner jet of Centaurus~A
\citet{hardcastle_03a} find a number of knots, which do not have any
detectable proper motion. Because they are almost stationary with respect to
the relativistic flow in the jet, they are either candidates for standing
shocks or shocks in the downstream flow of a collision with an obstacle
\citep{goodger_10a} -- both are capable of in-situ particle acceleration
within the jet. However, even in Centaurus~A there are strong indications for
another, more diffuse acceleration process in the outer $3.4~\rm kpc$ of the
jet, because the X-ray emission is truly diffuse and does not contain any
knots \citep{kataoka_06a,hardcastle_07a}. Such diffuse X-ray emission is seen
also in the jets of many other FR~I sources \citep[e.g.,
3C~66B;][]{hardcastle_01a}.

The high-resolution map in Fig.~\ref{fig:n3801_6cm_ABC-conf}, which has a
spatial resolution of 100~pc, shows several knots of emission in the jet, with
the most prominent one about $6~\rm arcsec$ away from the nucleus in the eastern
lobe and three smaller knots in the eastern jet closer to the nucleus. In the
western jet, there is an extended knot $4~\rm arcsec$ away from the nucleus and no
other visible knots further downstream. But comparing these results with the
spectral age map in Fig.~\ref{fig:spec_ageing}, we do not find any striking
similarities. We do not resolve in the spectral maps the inner knots within
$1.25~\rm kpc$ from the nucleus, because the nucleus confuses the emission,
but we can already see that the jet is hardly younger than the surrounding
plasma. Hence, we may be witnessing the influence of diffuse shock
acceleration on the CRe and thus the RC emission. We notice that where the
eastern jet bends from a south-eastern direction by $45\degr$ to a western
direction, most of the younger CRe in the eastern lobe are found. In the
western lobe, where no such change of direction is seen, there is also no such
prominent area of younger CRe. This favours the in-situ re-acceleration
scenario, because if backflow was the cause the eastern and western jet should
have a similar CRe age along the jet. At the position where the jet bends,
there is a bright blob of hot X-ray emitting gas. We hence conclude that the
CRe are accelerated by shock waves downstream of the contact discontinuity,
where shocks upstream heat the ISM to X-ray emitting temperatures, in
agreement with the observations of Centaurus~A by \citet{croston_09a}.
The similarity to
Centaurus~A extends also to the RC spectrum: the young inner lobes in
Centaurus~A do not show any spectral steepening up to observing frequencies of
$43~\rm GHz$, indicating a very young CRe population with a
spectral age of less than 1~Myr \citep{alvarez_00a}. \citet{clarke_92a} showed
that the spatially resolved radio spectral index between $1.6$ and $4.9$~GHz
has only little variation in both inner lobes. This can be explained again by
an effective cosmic-ray mixing process such as convective transport in the
backflow of the magnetized plasma.
\subsection{The evolution of the radio lobes}
\label{dis:evolution}
\begin{figure}
\includegraphics[width=1.0\hsize]{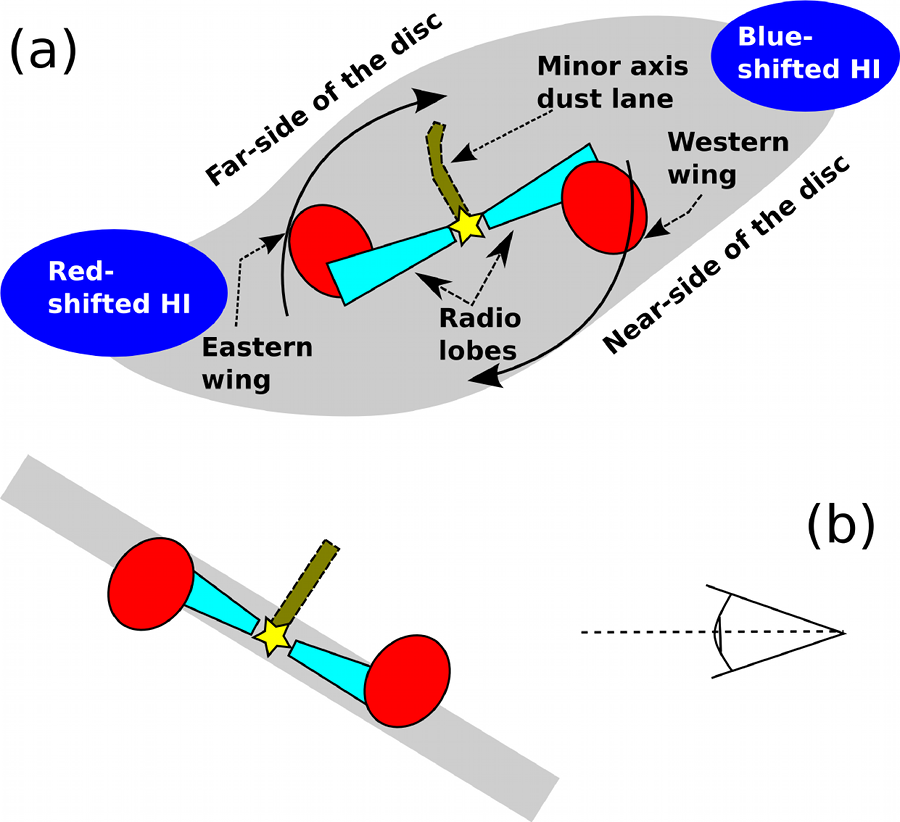}
\caption{Two possible viewing geometries of NGC~3801. In (a) the red bubbles
  that indicate the radio wings are downstream transported by the ISM. The
  hysteresis-like disc geometry is somewhat overstated for clarity. In (b) the
  radio wings expand along the vertical pressure gradient. (a) is a view as
  from the observer, while (b) is a view in profile from East. The minor axis
  dust lane is shown in green-brown in both panels.}
\label{fig:geometry}
\end{figure}

NGC~3801 is a young radio galaxy that we see in the early stages of its
eventual transition to become a more developed FR~I source. It has a so-called
`Z'-shaped symmetry, with the wings of the radio lobes pointing in opposite
directions. A possible explanation is that the jet hits a shell of denser gas
and gets deflected. The rotating gas in the ISM then moves the magnetized
plasma and shock heated gas downstream \citep{gopal-krishna_12a}. Also, the
radio lobes are more likely to expand along the vertical density gradient of
the ISM. Typical H\,{\small I} scale heights in spiral galaxies are only a few
hundred parsec which can be used as a reference, because much less is known
about the scale height in elliptical galaxies. It is difficult to establish
the exact geometry of the radio lobes, but we note that the jets are aligning
well with the major axis of the galaxy (Figs.~\ref{fig:high-res}a and b). As
noted by \citet{das_05a} it is very unlikely, however, that we are seeing the
galaxy in edge-on position and that the dust lane extending to the south-east
is indeed in the galactic plane. It is clear from H\,{\small I} measurements
that the east side is redshifted and the west side is blue shifted
\citep{hota_09a}. But we do not know whether the part north-east of the major
axis is on the far or near side. We can try to make an educated guess by
noting that the dust lane along the minor axis is much more prominent on the
north side of the nucleus. This dust is likely associated with the molecular
hydrogen as traced by the $^{12}{\rm CO}~J=1\rightarrow 0$ observations of
\citet{das_05a}. They found a gas ring with 2~kpc radius rotating around the
nucleus and aligned with the minor axis. Hence, the dust extends vertically in
front of the far side of the galaxy north-east of the nucleus, creating the
visible dust lane. In this scenario, the wings would be indeed downstream of
the jet explaining the `Z'-shaped symmetry. This geometry is shown in
  Fig.~\ref{fig:geometry}a ; for a five band multi-wavelength image, which
  encompasses the entire galaxy, we refer to \citet[][their
  fig.~1]{hota_12a}.
\citet{hota_09a} measured the circular rotation speed as ${\rm v}_{\rm
  circ}=270~\rm km\,s^{-1}$, which is a factor of three lower than the lobe
expansion speed of $850~\rm km\,s^{-1}$. We hence would expect the wings to
expand also into the direction opposite to the rotation of the gas, resulting
in a structure more reminiscent of the X-shaped lobes of 3C~305
\citep{hardcastle_12a}, which is not observed. Also, the fact that the
H\,{\small I} is observed in a ring surrounding the galaxy suggests that the
geometry is close to an edge-on position. The deflection of the lobes
downstream in the rotating ISM would be hardly visible because it is mainly
directed along the line-of-sight. The rotating ISM scenario may hence not
offer the sole explanation. An alternative scenario is that the jets lie not
exactly in the plane of the galaxy, but slightly inclined. In this case the
wings could expand vertically along the density gradient away from the gaseous
disc (Fig.~\ref{fig:geometry}b). Because the jets are in exactly opposite
directions, the wings would inflate into the opposite sides with respect to
the galactic plane, creating the `Z'-shaped geometry. Even if the jets do lie
exactly in the plane of the galaxy, any warping of the gaseous disc creates an
asymmetry. The stellar distribution indeed shows a pattern reminiscent of a
hysteresis loop \citep{heckman_86a,hota_12a}, indicative of such a warp. Thus,
we propose that for NGC~3801 the scenario of the rising wings along the
vertical density gradient of a gaseous, strongly warped, disc is the most
likely explanation for the `Z'-shaped symmetry.
\subsection{AGN feedback}
NGC~3801 has been proposed by \citet{hota_12a} as a galaxy that is in
transition from a star-forming galaxy to a so-called `red and dead'
galaxy. The extended 30~kpc ring of H\,{\small I} led \citet{hota_12a} to
conclude that NGC~3801 underwent a gaseous merger one billion years ago,
observing that the H\,{\small I} ring is very asymmetric having a
rotation period of 300~Myr \citep{hota_09a}. If the gaseous merger had happened
much longer ago, the H\,{\small I} distribution should be almost even. It is
likely that this gaseous merger has initiated the star formation and also
created the 2~kpc molecular ring that rotates perpendicular to the galactic
plane. \citet{hota_12a} found the presence of \emph{GALEX} FUV/NUV
emission, indicating that the star formation occurred in the last few $\times
100~\rm Myr$, whereas the young age (few Myr) of the radio lobes means that the
AGN has only recently started to become active. This is in accordance with the
results of \citet{schawinski_07a} who find that in general the AGN activity peaks about
500~Myr after the onset of the star formation.

In Figs.~\ref{fig:high-res}c and d we show an overlay of the RC
emission on to \emph{GALEX} FUV emission tracing star formation in the past
100~Myr. The distribution of the NUV/FUV emission is much more extended than
the size of the radio lobes, so that we are presenting here only the inner third of
the galaxy \citep[see][for the full map of FUV/NUV emission]{hota_12a}. The star formation is distributed in three main complexes
following the wisp-shaped dust lane. The radio lobes are situated in the gaps
between the star formation complexes, with the most prominent one roughly
aligned with the dust lane along the minor axis. The central
star forming complex is very similar to the disc of rotating molecular gas
suggesting an association. Clearly, the jets and the lobes have not
yet interacted with the molecular gas disc. The eastern lobe is surrounded by
a rim of FUV emission, with the lobe residing in a depression of FUV
emission. The wing of the western lobe is extending into a region, which
contains little FUV emission. Hence, the distributions of the RC and the
FUV emission display hints of an anti-correlation. It is possible
that we are witnessing the early stages of the interaction between the
star formation complexes and the radio lobes. Alternatively, the radio lobes
are just expanding into the areas where the gas density is lower as already
discussed in Section~\ref{dis:evolution}. It is sometimes proposed that
feedback by AGN lobes can possibly even trigger star formation
\citep[e.g.,][]{van_breugel_93a}. We see no indication for young stars of a
few Myr age, which would be detected by H~$\alpha$ emission \citep[see map by][]{verdoes_kleijn_99a}. On the other hand, the age of the radio lobes is so small
that it may well be that such stars are not yet formed or are still enshrouded by dust.
With the current expansion speed of the lobes the lobes will have expanded to
the size of the galaxy in $\approx 10~\rm Myr$, thus possibly disrupting the
star formation and eventually shutting it down.  Our preferred scenario is
hence that NGC~3801 is an example for a galaxy where the AGN feedback is able
to quench star formation and hence shape the galaxy luminosity function as
suggested by theoretical models \citep[e.g.,][]{croton_06a}.
\section{Conclusions}
We have observed the kpc-sized radio galaxy NGC~3801 with the VLA at
frequencies between 22 and 35~GHz and combined the new data with VLA and BIMA
archive data in order to study the RC emission between $1.4$ and $112.4~\rm
GHz$. We estimated the magnetic field strength by energy equipartition and
used pixel-by-pixel based fitting routines from the program {\small BRATS}
\citep[Broadband Radio Analysis ToolS,][]{harwood_13a} to fit JP model spectra
to the RC maps. Our main conclusions are:
\begin{enumerate}

\item The integrated RC flux density of the radio lobes has a spectral index
  break at $\approx 10~\rm GHz$. The spectral index steepens from $\alpha_{\rm
    low}=0.6\pm 0.1$ between $1.4$ and $4.9~\rm GHz$ to $\alpha_{\rm
    high}=1.6\pm 0.4$ between 22 and 35~GHz. A JP-model
  fit to the integrated data results in a spectral age of $\tau_{\rm
    int}=2.0\pm 0.4~\rm Myr$.

\item The dynamical age of the radio lobes is $\tau_{\rm dyn}=4~\rm Myr$,
  using the expansion velocity of the lobes as measured from X-ray
  observations by \citet{croston_07a}. The remarkably good agreement between
  spectral and dynamical age corroborates the scenario proposed by
  \citet{croston_07a}, where the hot X-ray emitting gas surrounding the radio
  lobes in shells is created by shock heating due to the expansion of the
  radio lobes. The energy as contained in the lobes is comparable to the
  energy of the ISM of the host galaxy, clearly able to influence its further
  evolution. NGC~3801 is a candidate for an elliptical galaxy, where in the
  aftermath of a gaseous merger star formation is to be suppressed in the next
  few 10~Myr due to feedback by its AGN.

\item The distribution of the radio spectral index across the radio lobes has
  only little variation, requiring an effective mechanism to re-distribute CRe
  within the lobes. A JP-model fit to the maps results in a map of the CRe
  spectral age, as derived by a superposition of mixed CRe populations. We
  find values between $1.8$ and $2.4~\rm Myr$, where the youngest CRe are
  found at the jet termination points and the oldest in the wings of the radio
  lobes. The small age variation requires a CRe bulk velocity of $4200~\rm
  km\,s^{-1}$, where the CRe are transported convectively within the backflow
  of the magnetized plasma surrounding the relativistic jet.

\item In-situ re-acceleration of CRe is happening at the jet termination
  points, both in the eastern and western radio lobe. The integrated radio
  spectral index between $22$ and $112.4~\rm GHz$ is with $\alpha=1.1\pm 0.2$
  flatter than expected for a pure JP model. This suggests ongoing CRe
  acceleration within the radio lobes. The spatial correlation between the hot
  X-ray emitting gas and the areas with a younger CRe spectral age
  means that the CRe are accelerated in shock waves downstream of the contact
  discontinuity, where the shocks upstream are responsible for heating the ISM
  to temperatures where they emit thermal X-ray emission. 

\item The JP-model fit allows us to measure the injection spectral index as
  $\alpha_{\rm inj}=0.5\pm 0.05$. This is in excellent agreement with
  first-order Fermi CRe acceleration by strong, non-relativistic shocks
  \citep{bell_78a,blandford_78a}.

\item The closest agreement between spectral and dynamical age is found for an
  equipartition magnetic field strength of $B=11~\rm\mu G$, with no
  contribution from protons, assuming a pure electron--positron
  plasma. NGC~3801 is a young radio galaxy, where entrainment of protons does
  play a lesser role than in more developed FR~I radio galaxies such as 3C~31
  \citep{croston_14a}.

\end{enumerate}

\section*{Acknowledgements}
{VH and JHC acknowledge support from the Science and Technology Facilities
  Council (STFC) under grant ST/J001600/1. JJH wishes to thank the STFC for a
  studentship and the University of Hertfordshire for their generous financial
  support. This research has made use of the NASA/IPAC Extragalactic Database
  (NED) which is operated by the Jet Propulsion Laboratory, California
  Institute of Technology, under contract with the National Aeronautics and
  Space Administration. The anonymous referee is thanked for several
  constructive comments that improved the clarity of the manuscript.}
\bibliographystyle{mn2e}
\bibliography{jets_bib}

This paper has been typeset from a \TeX/\LaTeX file prepared by the author.

\end{document}